\documentclass[10, journal, compsac]{IEEEtran}

\IEEEoverridecommandlockouts
\usepackage{amsmath,amssymb,amsfonts}
\usepackage{hyperref}
\usepackage{xspace}
\usepackage{color}
\usepackage{listings}
\usepackage{acronym}
\usepackage[T1]{fontenc}
\usepackage{lmodern}
\usepackage{multirow}
\usepackage{todonotes}
\usepackage{url}
\usepackage[export]{adjustbox}
\usepackage{tcolorbox}
\usepackage{comment}
\usepackage{colortbl}
\usepackage{multirow}
\usepackage{fancyhdr}
\usepackage{subcaption}%
\usepackage{graphicx}
\usepackage{algorithm}
\usepackage[utf8]{inputenc}
\usepackage{algpseudocode}
\usepackage{graphicx}
\usepackage{lipsum}
\usepackage{svg}
\setsvg{inkscape=inkscape -z -D} 
\definecolor{lightgray}{rgb}{.9,.9,.9}
\definecolor{darkgray}{rgb}{.4,.4,.4}
\definecolor{darkgreen}{rgb}{0, 0.39, 0.00}
\definecolor{Gray}{gray}{0.7}

\acrodef{CACC}{Cooperative Adaptive Cruise Control}
\acrodef{ECU}{Electronic Control Unit}
\acrodef{RSU}{Road Side Unit}
\acrodef{SC}{Service Center}
\acrodef{ITS}{Intelligent Transportation System}
\acrodef{OBU}{On-Board Unit}
\acrodef{OBD}{On-Board Diagnostics}
\acrodef{CAN}{Controller Area Network}
\acrodef{DoS}{Denial of Service}
\acrodef{AAA}{Automobile Association of America}
\acrodef{V2V}{Vehicle-to-Vehicle}
\acrodef{V2I}{Vehicle-to-Infrastructure}
\acrodef{PID}{Priority ID}
\acrodef{CIDS}{Clock-based IDS}
\acrodef{RLS}{Recursive Least Squares}
\acrodef{IDS}{Intrusion Detection System}
\acrodef{IVI}{In Vehicle Infotainment}
\acrodef{HMM}{Hidden Markov Model}
\acrodef{ML}{Machine Learning}
\acrodef{SAE}{Society of Automotive Engineers}
\acrodef{RPM}{Revolutions Per Minute}
\acrodef{CSMA/CD}{Carrier Sense Multiple Access with Collision Detection} 
\acrodef{ML}{Machine Learning}
\acrodef{NN}{Neural Network}
\acrodef{TPM}{Tire Pressure Monitoring System}
\acrodef{GPS}{Global Positioning System}
\acrodef{DOS}{Denial of Service}
\acrodef{DNN}{Deep Neural Network}
\acrodef{LSTM}{Long Short-Term Memory}
\acrodef{MSG}{Messages-Sequence Graph}
\acrodef{CPD}{Change Point Detection}
\acrodef{RNN}{Recurrent Neural Network}

\title{Detection of Message Injection Attacks onto the CAN Bus using Similarity of Successive Messages-Sequence Graphs}

\author{\IEEEauthorblockN{Mubark Jedh
\IEEEauthorrefmark{1}, Lotfi ben Othmane\IEEEauthorrefmark{1}~\IEEEmembership{IEEE Senior Member}, Noor Ahmed\IEEEauthorrefmark{2}, Bharat Bhargava\IEEEauthorrefmark{3}~\IEEEmembership{IEEE Fellow}}
\\
\IEEEauthorblockA{\IEEEauthorrefmark{1} Iowa State University, Ames, IA USA}%}
\IEEEauthorblockA{\IEEEauthorrefmark{3}Purdue University, West Lafayette, IN USA}%}
\IEEEauthorblockA{\IEEEauthorrefmark{2}Air Force Research Lab, Rome, NY USA}%}
}

\begin{document}
\maketitle

\begin{abstract}
The smart features of modern cars are enabled by a number of \acfp{ECU} components that communicate through an in-vehicle network, known as \ac{CAN} bus. The fundamental challenge is the security of the communication link where an attacker can inject messages (e.g., increase the speed) that may impact the safety of the driver. Developing an effective defensive security solution depends on the knowledge of the identity of the \acp{ECU}, which is proprietary information. This paper proposes a message injection attack detection mechanism that is independent of the IDs of the \acp{ECU}, which is achieved by capturing the patterns in the message sequences. First, we represent the sequencing of the messages in a given time-interval as a direct graph and compute the similarities of the successive graphs using the cosine similarity and Pearson correlation. Then, we apply threshold, change point detection, and \ac{LSTM}-\ac{RNN} to detect and predict malicious message injections into the CAN bus. The evaluation of the methods using a dataset collected from a moving vehicle under malicious RPM and speed reading message injections show a detection accuracy of 98.45\% when using \ac{LSTM}-\ac{RNN} and 97.32\% when using a threshold method. Further, the pace of detecting the change is fast for the case of injection of RPM reading messages but slow for the case of injection of speed readings messages. 

\end{abstract}

\section{Introduction}

\begin{figure}[tbp]
\centering
	\includegraphics[width=.45\textwidth]{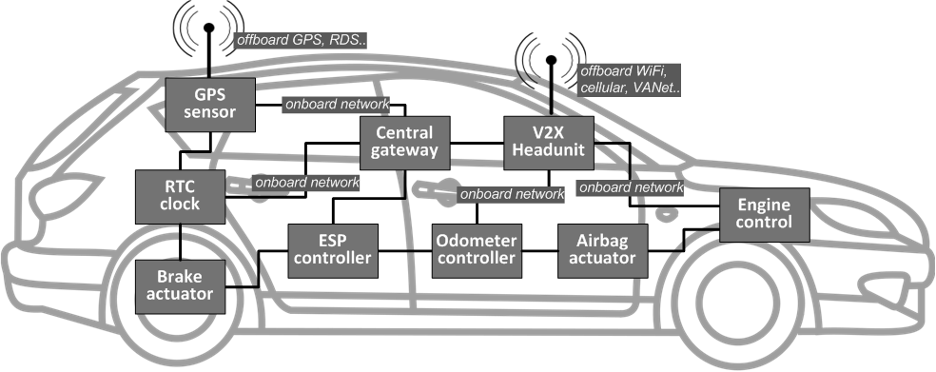} 
\caption{Example of connected vehicle~\cite{Othmane2015}}
\label{fig:connectvehicle}
\end{figure}

The growing market explosion on modern cars with high premium prices is driven by the increased consumer awareness for their safety features and superior functionalities. This is credited to a number of \acfp{ECU} components that communicate through an in-vehicle network, known as \acf{CAN} bus~\cite{bosh1991}. Within a single vehicle, there is a complex network of around one hundred collaborating \acp{ECU}, as depicted in Figure~\ref{fig:connectvehicle}. These \acp{ECU} use a large software base of about 100MB to control the functionalities of the vehicle through message exchanges using the CAN bus~\cite{4562233}. There are two types of \ac{CAN}: (1) CAN-H (high), a fast bus for communicating critical data, such as the engine, transmission, and speed messages, and (2) CAN-L (low), a slow bus for communicating non-critical data such as infotainment data~\cite{4562233}. Most importantly, it's used for safety mechanisms such as; collision avoidance, anti-lock brakes, traction control, and electron stability control. 

Furthermore, the CAN bus enables intercommunication link within the vehicle and external vehicles and devices through WiFi, Bluetooth, or cellular networks. This capability is exploited by the \ac{ITS} applications, such as infotainment systems, fleet management systems, parking assistance, remote diagnostics, eCall, remote engine start, and \ac{CACC} systems. These applications communicate with the \acp{ECU} of the vehicle through the \ac{CAN} bus to improve the experience of the customers~\cite{Othmane2015,HHZ2018,OABF2018} by, for example, reducing the speed and activating the brakes of the vehicle.

% \lipsum
  \begin{figure*}[tbp]
  \centering
  \begin{subfigure}[t]{0.4\textwidth}
        \includegraphics[width=1.0\linewidth]{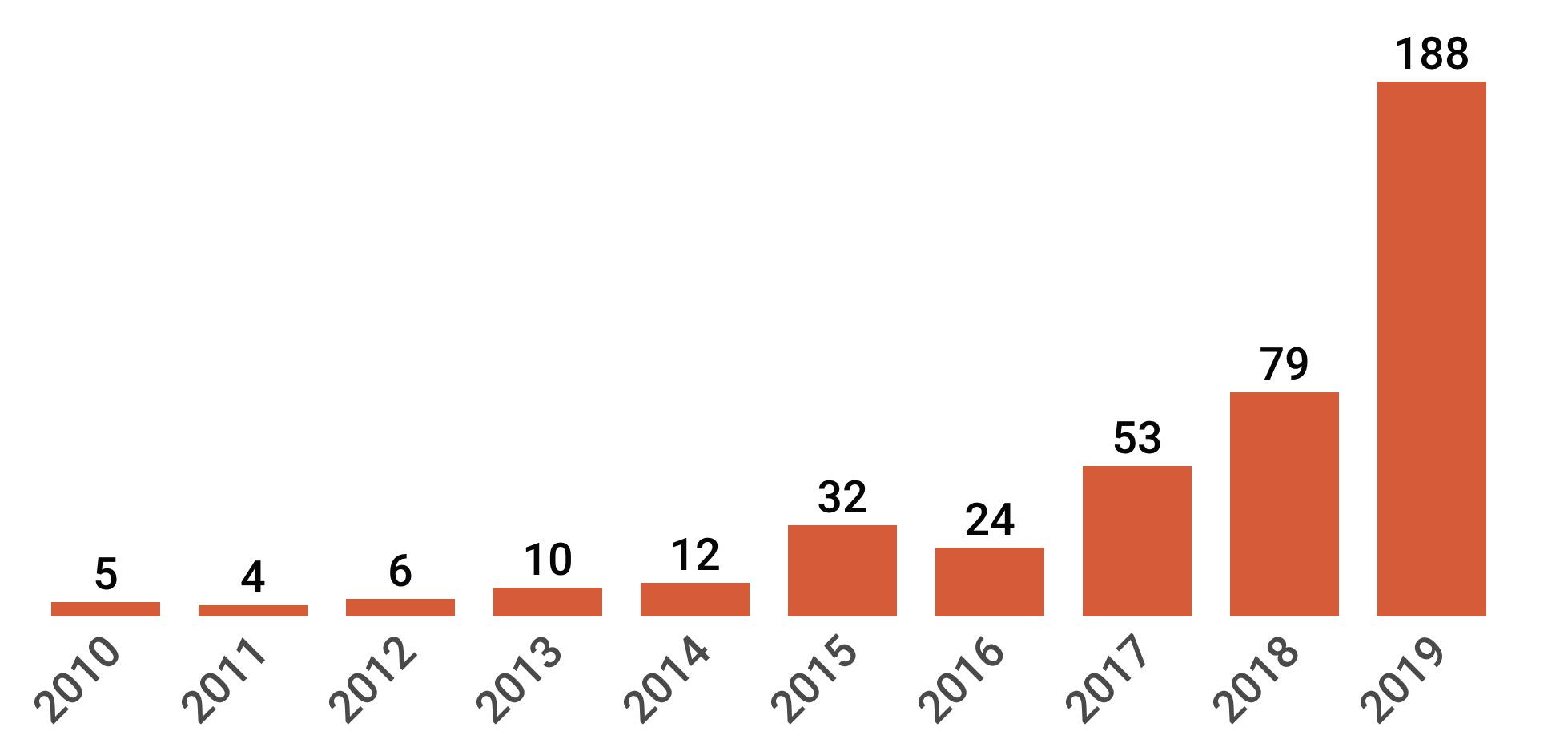}
  \caption{Increase of cyber-attacks on connected vehicles.}
  \label{fig:incidentgrowth}
  \end{subfigure} 
~
\begin{subfigure}[t]{0.4\textwidth}
        \includegraphics[width=1.0\linewidth]{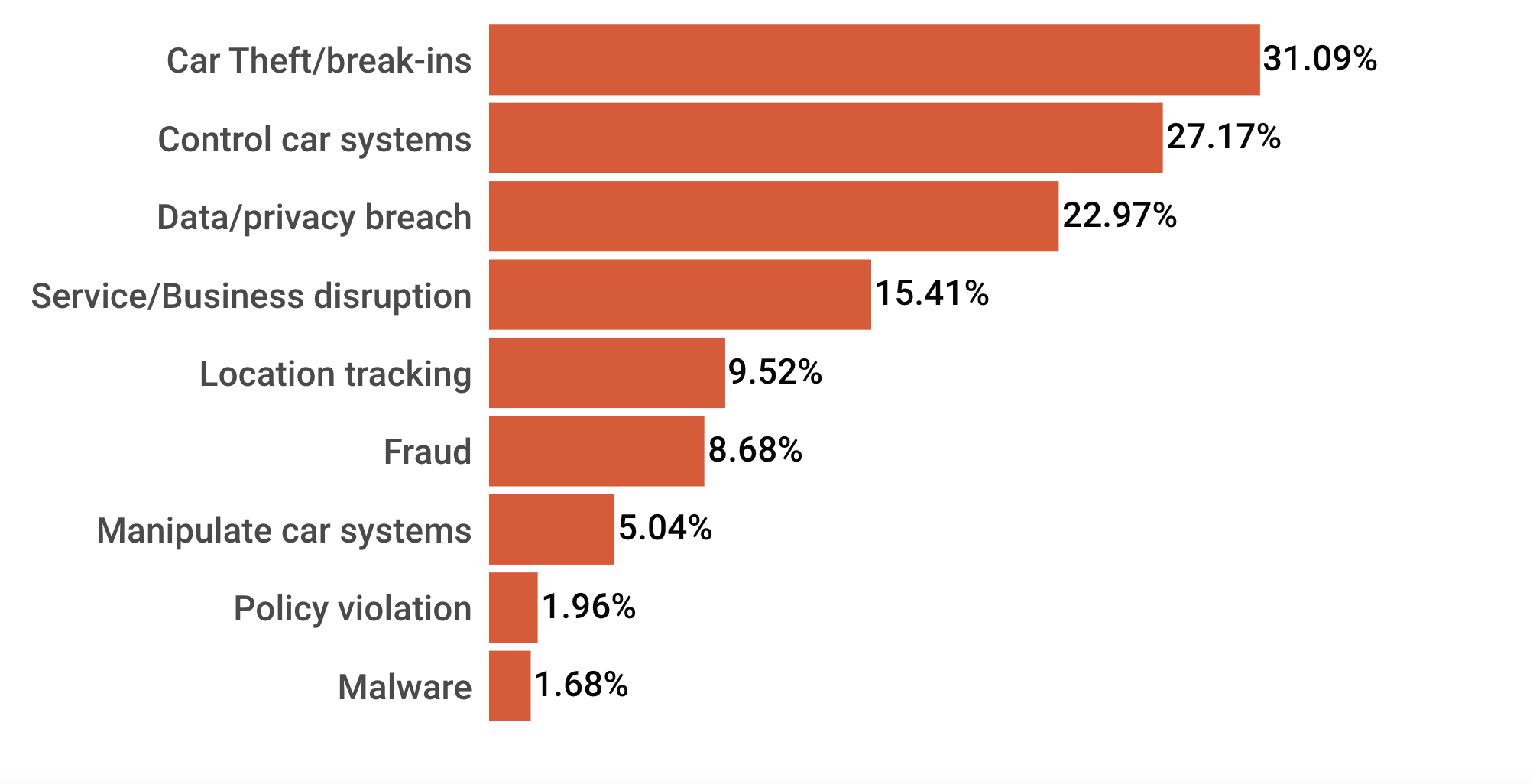}
  \caption{Impacts of cyber-attacks on connected vehicles.}
  \label{fig:incidentimpact}
  \end{subfigure}  
 % \hfill 
  \caption{Growth and distribution of cyber-attacks on connected vehicles between 2010 and 2019~\cite{UpstreamAuto2020}.}
  \end{figure*}

The CAN bus was designed as a stable, safe, and flexible closed network without considering security, specifically, authentication and authorization mechanisms. In addition, the extension of the network through the On-Board Diagnostics (OBD) to provide ways to report self-diagnosed errors and malfunctions through \acp{OBU} contributed to expanding the attack surface (the set of ways an attacker can compromise the vehicle) through the CAN bus~\cite{ValMill,Brandom,Golson}. Upstream’s research team identified 367 publicly reported incidents for a decade long~\cite{UpstreamAuto2020}. The analysis of these incidents shows an exponential growth of attacks, as depicted by Figure~\ref{fig:incidentgrowth}. Among these attacks, 27\% involved taking control of the car, as depicted by Figure~\ref{fig:incidentimpact}.

To remedy these security problems, a wide array of defensive security solution schemes has been proposed. Most of these solutions typically require the modification of the \ac{CAN} protocol, which is not practical for aftermarket vehicles~\cite{Othmane2015,HHZ2018}. Others have attempted to devise machine learning-based solutions driven by syntactic data through simulations or data related to researchers' devices to detect malicious behaviors~\cite{9076852}, which limits the practicality of their efficacy. Many of the practical solutions require the knowledge of CAN ID of the messages (the \ac{CAN} messages include an ID that indicates the information embedded in the message, e.g., speed and brake) which is proprietary to the car manufacturers~\cite{SGL2019}. 

In this paper, we answer the question: {\it Can we detect cyber-attacks on a moving vehicle without the need to know the IDs of the vehicle's \acp{ECU}?}
We hypothesize that there is a pattern of messages sequences between the collaborating \acp{ECU} of the vehicle, and injecting CAN messages disrupts the sequences' pattern. To answer this question, we first develop a direct graph that represents the sequence relations between \acp{ECU} messages based on their CAN ID, which we call \emph{\ac{MSG}}. Then, we use a sliding-window approach to compare the similarity between the graphs computed from the sequences of messages sent through the \ac{CAN} bus in successive time windows using the \textit{cosine similarity} and \textit{Pearson correlation} metrics. Next, we apply the \textit{threshold, change point detection, and \ac{LSTM}-\ac{RNN}} techniques to detect and predict injection attacks on the CAN bus.  
The main contributions of the paper are:
\begin{enumerate}
    \item A graph-based similarity metrics that capture the variations in CAN messages patterns caused by messages injection attacks on the CAN Bus. 
    \item An effective \ac{LSTM}-\ac{RNN}-based \ac{ML} technique for detecting messages injections attacks on CAN Bus from \acp{MSG}.
    \item A change-point detection technique for detecting messages injection attacks on CAN Bus from \acp{MSG}.
\end{enumerate}

The paper is organized as follows. Section~\ref{sec:background} discusses the security issues of the \ac{CAN} protocol and Section~\ref{sec:relwork} reports about related work. Then, Section~\ref{sec:approach} presents the research method, Section~\ref{sec:experimental} describes the evaluation methods and Section~\ref{sec:results} reports the results. Section~\ref{sec:conclusions} concludes the paper.

\section {Security Issues on CAN Protocol}\label{sec:background}

The \acf{CAN} is a network protocol developed by Robert Bosch in 1986~\cite{bosh1991} to communicate the \acp{ECU} that control the behaviors of the vehicles' mechanical and electrical components~\cite{Embedded-system}. All cars manufactured in USA after 2008 support the \ac{CAN} protocol and offer a means to interface with it through the \acf{OBD}-II port, which is usually located under the steering wheel. The OBD is implemented in the vehicles to provide ways to report self-diagnosed errors and malfunctions~\cite{OBDII} and is mandatory for all cars and light trucks sold in the United States and the European Union. This port gives direct access to the CAN network, which inherently creates the attack surface, the set of ways an attacker can penetrate the vehicle.

By design, CAN bus is resilient, robust, and easy to wire but has a set of security weaknesses. The fundamental security issues are: (1) The CAN bus data frame has no source identifier field to identify the legitimacy of the sending \ac{ECU}. As a result, \acp{ECU} cannot trust messages based on their sources. (2) The protocol does not protect the confidentiality of CAN messages. Attackers can read the messages exchanged in the CAN bus and infer information that could be used to stage attacks, including associating \acp{PID} with \acp{ECU}. (3) The use of priority in gaining the right to send messages could be easily used to flood the bus of a vehicle with messages that have small \acp{PID} and prevent other \acp{ECU} from communicating and makes their service unavailable, i.e., \ac{DOS} attack. frame~\cite{DBLP:journals/corr/abs-1802-01725}.

Typically, attackers inject messages into the \ac{CAN} bus, suppress legitimate messages (i.e., legitimate messages that have higher priority IDs than injected messages are ignored) or compromise an ECU to behave maliciously through either direct or indirect access points. Direct access point attacks include the OBD-II port, the CD player or the USB port~\cite{article1}. For instance, the attacker can inject malicious CAN messages through the USB port, encode a malicious software onto a CD to exploit the entertainment system of a car or use it to access the CAN bus or update the firmware of an ECU. The indirect access points are short and long-distance wireless access points connected to the \ac{CAN} bus. These include Bluetooth, on broad Wi-Fi such as \ac{V2V} devices, remote keyless entry, \ac{TPM}, \ac{GPS} and cell phone network. 

The attack surface increases as more features, such as vehicle apps used for telematics services, road-side assistance, \ac{V2V} applications, and remote diagnostic, are added in the car. Connecting vehicles to external entities through wireless devices expose the security weaknesses of the CAN bus~\cite{Othmane2015}.
Wolf et al. were among the pioneers in describing the weaknesses of the CAN bus, including unauthorized access into the CAN bus and lack of confidentiality and integrity checks of CAN messages~\cite{article}. Further, Hartzell et al. discussed the impacts of the common entry points, limited bandwidth, multi-cast messaging, lack of encryption of messages, and multi-system integration on the security of the CAN bus~\cite{article1}. 

Valeseka and Miller demonstrated that hijacking connected vehicles is possible. They identified remotely the IP address of their Jeep car from the Sprint Cellular network and took over the vehicle's critical features, including disabling breaks, controlling the steering wheels, and turning on/off windshield wipers while the car is moving~\cite{25}. Recently, Golston and Green reported successful exploits of vulnerabilities to take control of recent Tesla cars~\cite{golson2016,green2020}. An experimental security analysis of the attack surface of the connected vehicle including the short and long-range wireless channels, the entertainment systems (i.e., CD player, iPod port, USB port) and the electric charge port was performed by Koscher et al.~\cite{5504804}, who showed the wide range of potential attacks on connected vehicles. Othmane et al.~\cite{OFRB2014} surveyed security experts and validated Koscher et al.'s insights that attacks on connected vehicles are not lab experiments anymore. Thus, an effective and practical detection and prediction of injections of CAN messages is critical for connected and autonomous vehicles.

\section{Related work}\label{sec:relwork}

Attackers take control of connected vehicles by injecting messages into their in-vehicle networks, mainly their CAN bus. Several hardware and software-based encryption methods were proposed to prevent eavesdropping CAN messages. For instance, Farag et al. proposed CANTrack, which encrypts the data payload field of CAN messages to prevent access to them by non-legitimate entities~\cite{7934878}. On the one hand, message authentication mechanisms are considered to detect and/or prevent injection attacks. For instance, Hiroshi et al. proposed the use of a secret key that would be distributed to legitimate ECUs of the given vehicle~\cite{28}. Each of the legitimate \acp{ECU} of the vehicle must reject CAN messages that are not authenticated with the key. Authentication-based mitigation solutions have been heavily explored~\cite{inproceedings,10.1007/978-3-642-35404-5_15,inproceedings2,inproceedings3}, but they cannot be used for aftermarket vehicles. 

Other notable mitigation techniques have been reported in e.g.,~\cite{HHZ2018, Othmane2015}. In general, some of these proposed mitigation solutions trade performance and security requirements given the computation and communication constraints of the \ac{ECU} used in vehicles and time-criticality of some of the messages (i.e., braking). Others, require modification of the currently well-vetted CAN bus protocol (with respect to safety) which makes them not practical, especially for aftermarket vehicles. 

Early detection of message injection attacks on CAN bus methods were based on frequency, entropy, and correlation~\cite{article3}. For instance, Taylor et al. proposed an \ac{IDS} based on messages frequency using the Hamming distance between successive message data fields to detect attacks~\cite{7420322}. The methods were validated using syntactic dataset, a data set that is constructed by adding and removing messages to the set of the CAN messages collected in normal driving conditions. \acp{IDS} that are based on frequency, timing, and entropy are effective for only attacks that disturb the frequencies of CAN bus messages~\cite{9076852} because they assess the collision management in the \ac{CSMA/CD} algorithm used by the CAN bus.

Recently, there has been increased attention for the use of machine learning techniques based on feature extraction from the CAN bus data frame messages to identify attacks. At the core of these solution methods is discriminating messages associated to attacks and those that are not, with acceptable accuracy and false positive~\cite{10.1371/journal.pone.0155781}. \ac{NN} has been the commonly used ML-based approach for designing \acp{IDS} for the CAN bus, e.g.,~\cite{Lokman2019, 8687274, 9262960, SGL2019}. The main problem with this supervised learning method is that it requires extensive time to develop high-performance IDS models from labeled data~\cite{SGL2019} and the trained models are specific for vehicle' make and model, and driver. The unsupervised ML-based \acp{IDS} use mostly message-timing, message-frequencies, and message-latency~\cite{SHH2018, 8688625}. These methods identify the CAN bus fuzzing through the assessment of the contention algorithm used in the CSMA/CD protocol. 

Islam et al. investigated the use of patterns of sequences of CAN messages types to detect attacks~\cite{IRYM2020}. They developed graph-based anomaly detection technique that models the sequence of messages exchanged in the CAN bus and used the chi-square metric to compare the median between the successive graphs and showed its efficacy. The method is known to have low power/efficiency (the probability to avoid wrongly failing to reject the null hypothesis) for moderate to large sample sizes.

\begin{figure*}
  \includegraphics[width=\textwidth]{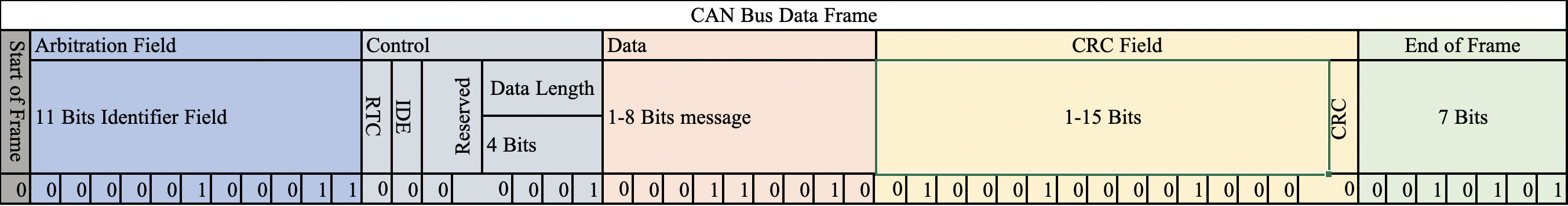}
  \caption{\ac{CAN} Message Data Frame Format}
  \label{fig:CANFrame}
\end{figure*}

\begin{figure}[tbp]
     \begin{subfigure}[t]{0.5\columnwidth}
        \includegraphics[width=1.0\linewidth]{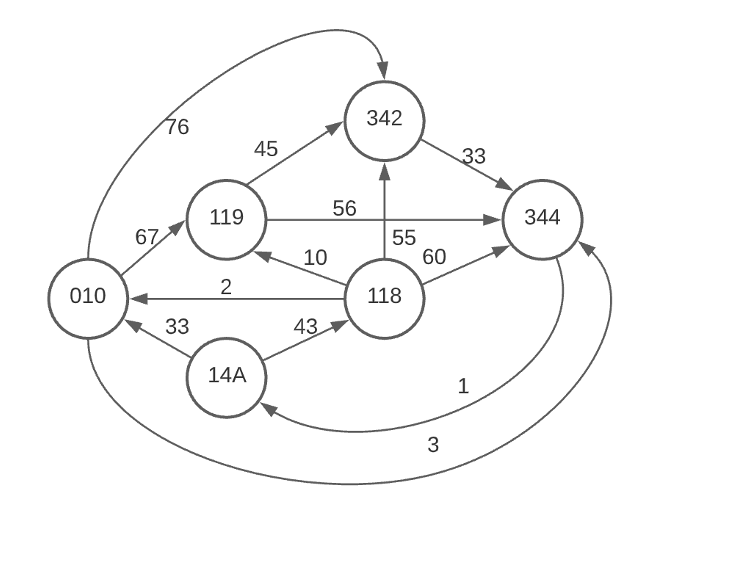}
        %\caption{MSG at time-window $t$.}
    \end{subfigure}%
     \begin{subfigure}[t]{0.5\columnwidth}
        \includegraphics[width=1.0\linewidth]{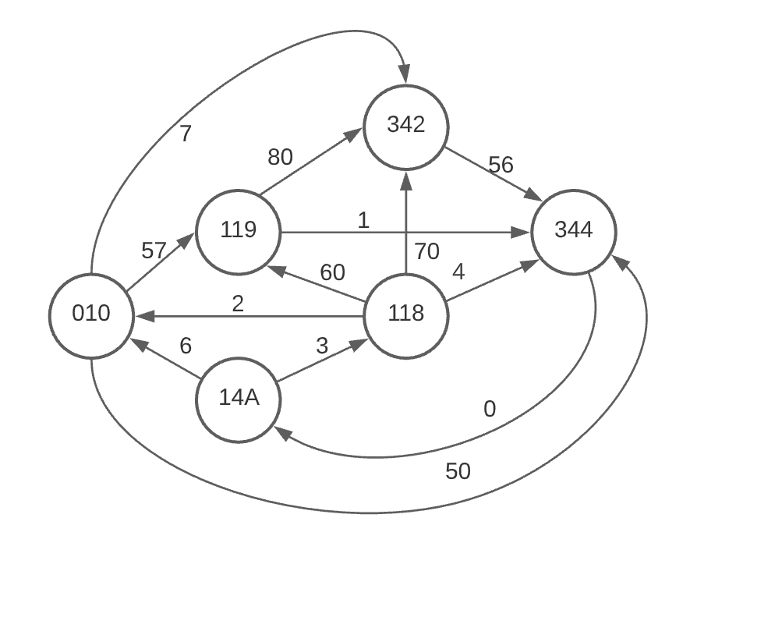}
        %\caption{MSG at time-window $t+1$.}
    \end{subfigure}%
     \caption{Similarity of two messages-sequence graphs at successive time-window $t$ (left) and $t+1$ (right). The labels of the nodes are the PIDs and the labels of the edges are number of times a message with the PID source of the edge is followed by a message with the PID destination of the same edge during the time-window. For example, 33 messages and 56 messages with ID 344 followed messages with ID 342 at resp. time-windows $t$ and $t+1$, which indicate a possible change of the behavior of the vehicle.
     }
    \label{fig:messagesgraph}
\end{figure}

The common data sources that are used to assess the accuracy of the ML-based IDS solution include use of data of own devices connected to a stationary vehicle, e.g.,~\cite{ChSh2016}, use of synthetic (artificial) data, e.g., ~\cite{TLJ20168}, use of data collected from simulators, e.g., ~\cite{LMKA2017}, and use of data from stationary vehicle, e.g., ~\cite{SHH2018, 9235336, IRYM2020}. This limits the confidence in the results and threatens its validity~\cite{CROT2016}. The evaluation of our IDS solutions uses datasets collected from an in-motion vehicle under injection of fabricated messages~\cite{othmane2020b}. This exceptions to this is the work of Stachowski~\cite{SGL2019} and Othmane et al.~\cite{9076852}, who used datasets collected from in-motion vehicles under messages injection attacks. Both studies require knowing the association of CAN ID to the vehicle's ECUs, which is proprietary information and dependent on the make or model.

Wu et al.~\cite{8688625} and Young et al.~\cite{8640808} provide comprehensive surveys on \ac{IDS} for connected and autonomous cars. Unlike most machine learning-based studies in the literature, our proposed technique neither depends on the make or model of the car nor its proprietary information (i.e., CAN ID); it uses a combination of graph-based and machine learning techniques.

\section{Research Method}\label{sec:approach}

\begin{figure}[tb]
\centering
 \includegraphics[width=0.48\textwidth]{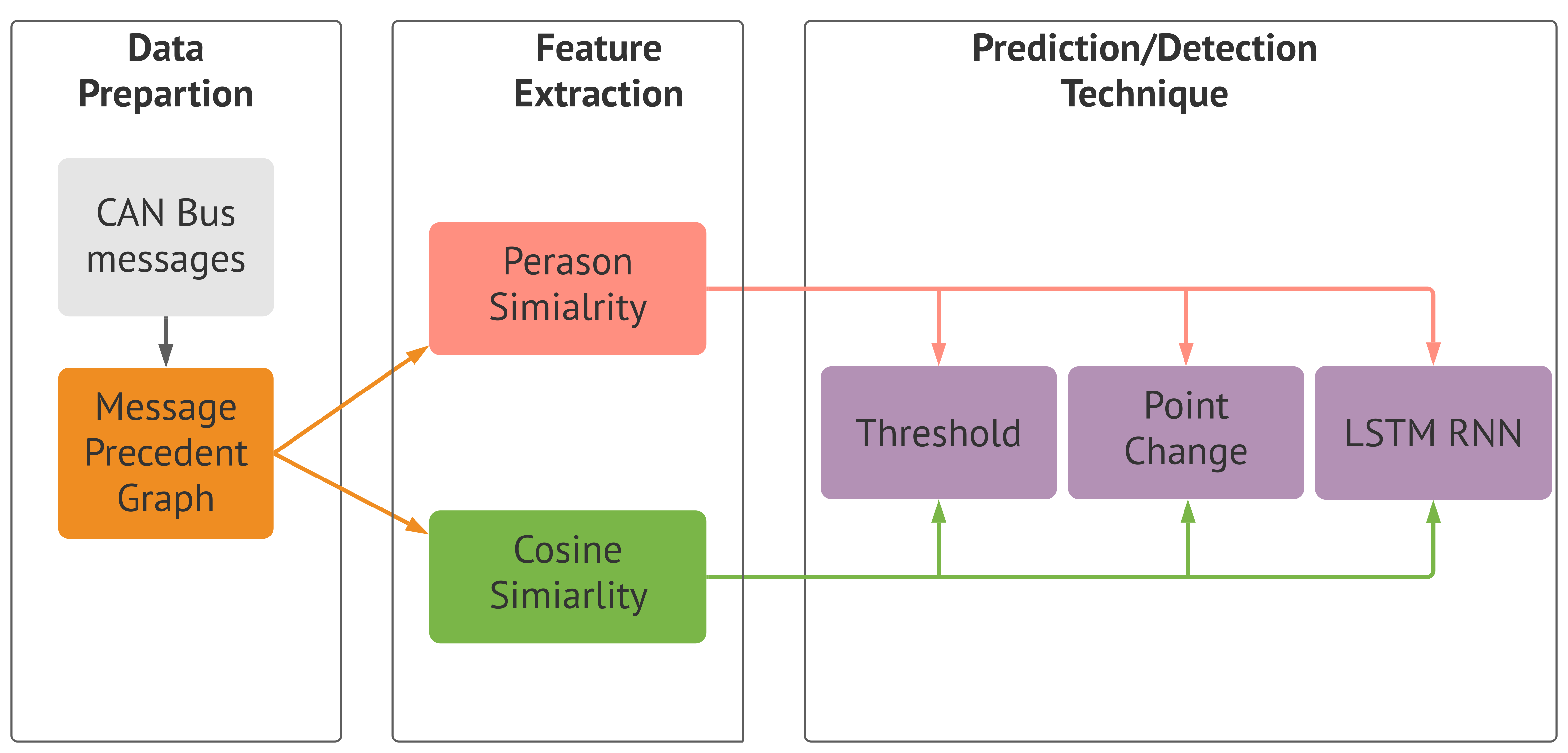}
   % \includesvg[width=1.1\columnwidth] {Figures/ResearchApproach.svg}
  \caption{High-level View of Our Solution Methodology}
  \label{fig:activity_diagram}
\end{figure}

We approached the problem of detecting and predicting message injection attacks on modern cars by capturing the patterns of the sequencing of the \ac{CAN} messages exchanged among the \acp{ECU}. As depicted by Figure~\ref{fig:activity_diagram}, data collected from the CAN bus are represented as a direct \textit{Message-Precedence} graph (left box). Then, the \textit{Pearson and Cosine} similarities of successive graphs are used as the \ac{ML} features (middle box). Finally, \textit{LSTM-RNN, Threshold, and Point Change} prediction is used to predict messages injection (right box). We validate our methodology using datasets collected from a moving car while fabricated speed and RPM messages are being injected into it's CAN bus. We describe these methods in the following subsections.

\subsection{Message-sequence graph}
Technically, CAN is a time-synchronized broadcast, multi-cast reception message network bus. Figure~\ref{fig:CANFrame} above illustrates the CAN message format. Each message consists of a data frame, remote frame, error frame, or overload frame. The data frame is used to exchange data between the nodes, the remote frame is used to request the transmission of a specific identifier, the error frame is transmitted by any node detecting an error, and the overload frame is used to inject a delay between data or remote frames~\cite{bosh1991}. 
\begin{figure}[H]
    \centering
    \includegraphics[width=0.48\textwidth]{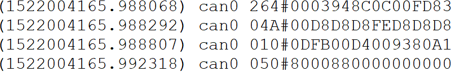}
    \caption{An Illustration of \ac{CAN} data stream format-- Each row corresponds a Timestamp followed by the \ac{CAN} bus ID "can0", the process \ac{PID} of the \ac{ECU} (i.e., 264), followed by "\#" and the actual data in hexadecimal format.}
    \label{fig:datasteam}
\end{figure}

Figure~\ref{fig:datasteam} shows a snapshot of a message captured from the CAN bus. Each message has a freshness tag (like a timestamp) followed by the \ac{CAN} bus ID, the \ac{CAN} ID of the \ac{ECU}, and the actual data. The \ac{ECU} of a given vehicle collaborate by exchanging successive messages through the CAN bus to perform specific actions. For example, a CAN message representing an increase of the fuel will usually be followed by a message representing an increase of RPM and most likely a message representing an increase of the speed~\cite{9076852}. It is intuitive to see that each of these messages have a sequence pattern; hence,  the core of our solution for detecting injection attacks.

With this, \textit{we hypothesize that there are patterns of sequences of messages exchanged between the collaborating \acp{ECU} of the vehicle and injecting CAN messages disrupts the messages-sequence patterns}. The idea is to capture the pattern of the sequences of the messages and represent it as a graph where the nodes represent the \acp{PID} and the edges represent the sequences of the messages of that given \acp{PID}. Note that the messages captured are independent of vehicle's \acp{ECU}.

Formally, let $E(N_1,N_2)$ be an edge that represents a CAN message with PID $N_1$ followed by a CAN message with PID $N_2$. Let the label of edge  $E(N_1,N_2)$ represents the frequency of CAN messages with PID $N_1$ followed by CAN messages with PID $N_2$. We call this graph \emph{\acf{MSG}} as shown in Figure~\ref{fig:messagesgraph}, and the construction of the graph sequence is illustrated in Algorithm 1.

In Algorithm~\ref{alg:PIDsequencegraph}, we first create a dictionary of the \acp{PID} exchanged in the \ac{CAN} bus--The \acp{PID} become the labels of the rows and columns on a matrix representing the edges of the \ac{MSG} in line 2. Then, it loops over all the CAN messages that were exchanged during the time window $w$ (we use windows of e.g., 100 successive messages) and increases the label of the edge linking the node representing the \ac{PID} of a given message to the node repressing the \ac{PID} of the previously processed message in lines 3-7. Equation~\ref{eq:disribution} represents the distribution of the messages-sequences at time $t$.
\begin{equation}\label{eq:disribution}
    D(t) ={E(N_i,N_j)(t)}
\end{equation}
\begin{algorithm}[!t]
\caption{Message Graph Sequencer}
\label{alg:PIDsequencegraph}
\begin{algorithmic}[1]
\Require{$CANData$: A batch of CAN messages with size window}
\Require{$window$: Size of the batch messages}
\Ensure{$MSG$: message precedence graph}
\Statex
\Function{ComputeMSG}{$data$,$window$}
  \State {Call CreateDictionary()}

    \For{$k \gets 1$ to $N$}                    
            \State {key $\gets$ {concat(CANData[$j$]['PID']}
            \State {+ CANData[$j$+1]['PID'])}}
            
            \State {{MSG[key]} $\gets$ {MSG[key] +1} }
    \EndFor
    \State \Return {$MSG$}
\EndFunction
\end{algorithmic}
\end{algorithm}
\subsection{Feature extraction}

We hypothesised that the \ac{MSG} representing the messages exchanged in a CAN bus during a time slot $t$ with size $w$ is $similar$ to the \ac{MSG} representing the CAN messages exchanged during the following time slot $t+1$ with same size $w$ in the case of normal driving behavior and that injection of messages into the CAN bus disrupts this pattern. We formulate the \textit{Similarity} concept for our \ac{IDS} using Equation~\ref{eq:Similarity}. That is, the similarity \textit{Sim} at time \textit{t+1} is the similarity of the distributions of the messages-sequences $D$ at time \textit{t} and at time \textit{t+1}. 

\begin{equation}\label{eq:Similarity}
    Sim(t+1) = Similarity(D(t),D(t+1))
\end{equation}

There are several similarity metrics that measure the similarity between two graphs~\cite{Neman2010} including Cosine similarity, Pearson and Cramer correlation, chi-squared, T-test and Levene's tests. We briefly describe the two methods that we selected for our study and the rationale behind our choice. 

\noindent{\bf Cosine similarity} measures the cosine angle between two non-zero vectors and determines whether the angle point is in the same direction or not. It helps to tease apart the types and relationships of vertices in social networks~\cite{Neman2010}. Simply, it's metric measures the angle between two vectors as formulated by Equation~\ref{eq:cossim}, where $x$ and $y$ are two vectors. Cosine Similarity is often used to measure document similarity, mainly for plagiarism detection. Simply, the metric compares two documents by measuring the similarity of the vectors of the frequencies of the words in both documents, then outputs a number between 0 and 1. The closer the number is to 1, the more similar the two vectors.

\begin{equation}\label{eq:cossim}
  Cosim(x,y) = \frac{\sum_{i=1}^{n} (x_i \times y_i)}{\sqrt{\sum_{i=1}^{n} x_i^2} \times \sqrt{\sum_{i=1}^{n} y_i^2}} 	
\end{equation}

The cosine similarity for the two MSGs of Figure~\ref{fig:messagesgraph} is 0.66.

\noindent{\bf Pearson correlation.} The correlation measures the strength of the linear association between two non-zero vectors and is given by Equation~\ref{eq:pearson}. It is widely used to measure similarity~\cite{Neman2010}. It compares the similarity and returns values between 1 and -1. The closer the value to 1 or -1, the strong the modeled relationship is.

\begin{equation}\label{eq:pearson}
  PC(x,y) = \frac{\sum_{i=1}^{n} (x_i -\overline{x}) (y_i - \overline{y})}{\sqrt{\sum_{i=1}^{n} (x_i -\overline{x})}\sqrt{\sum_{i=1}^{n}(y_i - \overline{y})}}
  \end{equation}

The Pearson correlation coefficients of the two MSGs of Figure~\ref{fig:messagesgraph} is 0.63.

 \begin{figure*}[tbph]
    \centering

    \begin{subfigure}{0.33\linewidth}
        \centering
       \includegraphics[width=1.0\linewidth]{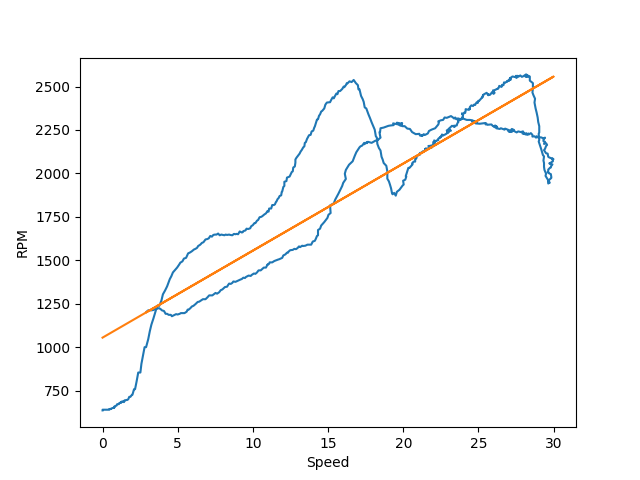}
        %\caption{No fabricated messages attack. In this drive test, the speed and RPM reached resp. almost 30 mph and 2300 units. The blue line shows the correlation between the speed readings and RPM readings.\\ \\ \\}
    \end{subfigure}%
              ~ 
    \begin{subfigure}{0.33\linewidth}
        \centering
       \includegraphics[width=1.0\linewidth]{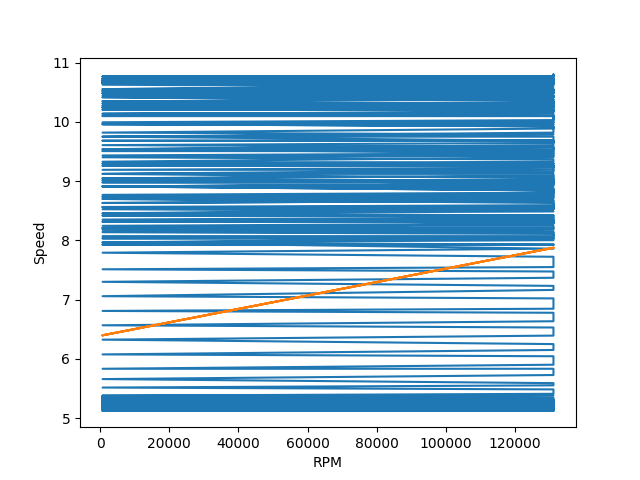}
        %\caption{Injection of fabricated RPM reading messages. The speed of the vehicle in this drive test reached almost 11 mph. The RPM readings oscillates between the actual values and the value that corresponds to the injected hex value "FFFF", i.e., 120000. The blue line shows the correlation between the speed readings and RPM readings in this experiment.}
    \end{subfigure}%
    ~ 
    \begin{subfigure}{0.33\linewidth}
        \centering
       \includegraphics[width=1.0\linewidth]{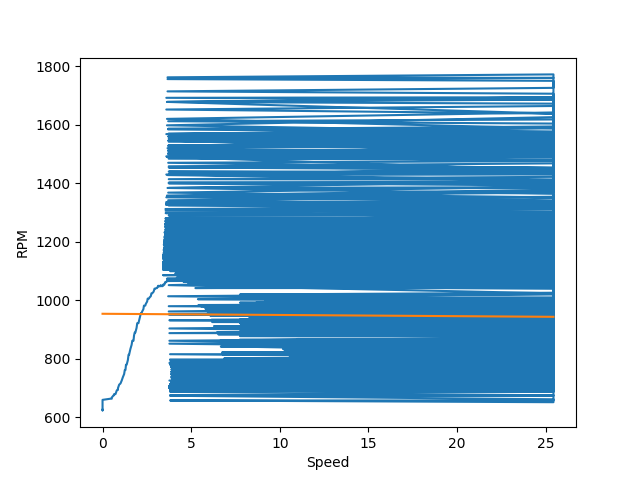}
        %\caption{Injection of fabricated speed reading messages. The RPM of the vehicle in this drive test reached almost 1800 units. The speed readings oscillates between the actual values and the value that corresponds to the injected hex value "FFF", i.e. almost 25 mph. The blue horizontal line shows that correlation between the speed readings and RPM readings is almost zero.}
    \end{subfigure}%
    \caption{An illustration of \textit{speed} and \textit{RPM} readings on the three datasets for a normal vehicle operation (left), and the impact on message injection attacks for RPM reading on speed reading (center), and on speed reading on RPM reading (right).}
    \label{fig:speedvsRPM}
\end{figure*}

\subsection{Prediction of messages injections} 
In conjunction with \textit{Cosine and Pearson} similarity metrics, we employed three techniques to predict the messages injection attacks, which are: threshold, \ac{LSTM}-\ac{RNN} and \ac{CPD}. In general, the threshold is commonly considered as the base case, and the \ac{LSTM} and \ac{CPD} are commonly used IDS solutions~\cite{TJS2018}.

\noindent{\bf Threshold.} A threshold is a selected value of the given similarity metric that is believed to provide better accuracy in detecting injection of messages. To identify "good" thresholds for the three selected similarity metrics, we variate the threshold values for the given metric and compute its accuracy until we observe an "optimal" value.

\noindent{\bf \acl{RNN}-\acl{LSTM}.} The technique was first introduced in 1997 and become the core methodology deep learning~\cite{279181}. It is a gradient-based architecture developed for modeling time-series data with long-term dependencies~\cite{HoSc1997}. The design solves the problem of vanishing gradient by allowing errors to be back-propagated through time. The \ac{LSTM} was constructed using the sequential model of a linear stack of layers, which are recurrently repeating memory blocks. It is very powerful in sequence prediction problems because it can store past information for a long time.

\noindent{\bf Change point detection.} \acf{CPD} is commonly used for detecting anomalous behavior from time-series data~\cite{OEMG2019}. The change point is defined as the point at which the parameters that describe time-series data e.g., mean and variance, abruptly change~\cite{BaHa1993}.

\section{Evaluation methods}\label{sec:experimental}

In this section, we discuss our experimental evaluation setup and the parameters used for our study. 

\subsection{Evaluation datasets}

In our previous study ~\cite{9076852}, we collected a log of CAN bus messages for (1) normal driving behavior, (2) injection of fabricated speed reading messages onto the CAN bus, and (3) injection of fabricated RPM reading messages onto the CAN bus of an in-motion Ford Transit 500 2017. The data set is available in~\cite{othmane2020b}. Table \ref{tab:listdatasets} shows the number of CAN messages that were used in the research. Figure~\ref{fig:speedvsRPM} shows the speed and RPM readings and their relationships in the three datasets~\cite{9076852}. The plot at the left shows that the speed of the vehicle varies between 0 and almost 30 mph and the RPM readings varies from almost 0 to almost 2300 units during a normal driving scenario. The plot at the center shows that the speed of the vehicle in this drive test reached almost 11 mph and the RPM readings oscillates between the actual values and the value that corresponds to the injected hex value "FFFF", i.e., 120000.  The plot at the right shows the RPM readings of the vehicle in this drive test reached almost 1800 units and the speed readings oscillates between the actual values and the value that corresponds to the injected hex value "FFF", i.e. almost 25 mph. 

The Pearson correlation between the speed readings and RPM readings represented by the orange lines in Figure~\ref{fig:speedvsRPM} shows a strong relationship (the coefficient is 0.85) between the 2 quantities in the plot at the left, weak relationship  (the coefficient is 0.33) in the plot at the center, and no relationship (the coefficient is -0.013) in the plot at the right~\cite{9076852}. This suggests that the \acp{ECU} of the vehicle may act on the injected RPM readings but may not act on the injected speed readings that they receive. The difference between the correlation coefficients hints to the impact of messages injection on the collaboration between the \acp{ECU} of the car, which we explore in this paper. 

\begin{table}[tbp]
\caption{Dataset Size}
\label{tab:listdatasets}
\centering
\begin{tabular}{p{0.1in} p{2.3in}p{.6in}}
\hline
\rowcolor{Gray}\hline
No &  Description & \# of \ac{CAN} messages \\\hline
\hline
  1   &  CAN Data for no injection of fabricated messages  & 23,963 \\\hline
  \rowcolor{lightgray}
  2   &  CAN Data with injection of "FFF" as the speed reading&  88,492 \\\hline
 3& CAN Data with injection of "FFFF" as the RPM reading & 30,308\\\hline
\hline
\end{tabular}
\end{table}

\subsection{Evaluation method for the similarity techniques}

We first construct the \acp{MSG} from each time-window (default 100) successive messages from the three CAN log datasets described in Table~\ref{tab:listdatasets}. Then, the similarity metrics are used to compute the similarity values of the \acp{MSG} series for each of the datasets.  

To assess the efficacy of each of the similarity metric, we plot the similarity values as time-series data to observe the tendencies of the values computed from the three datasets. Then, we plot the distributions of the frequencies of the similarity values to observe whether similarity values computed from the no injection of fabricated messages dataset differ or not from the similarity values computed from the RPM and speed injection of fabricated messages datasets. Next, we use t-test~\cite{Gosset1908} to statistically validate the difference between the two distributions of similarity values computed from the no injection of fabricated messages dataset and the distributions of similarity values computed from respectively RPM and speed injection of fabricated messages datasets.

In addition, we set thresholds for each of the similarity metrics and compared the similarity values computed from the \acp{MSG} to these thresholds. A similarity value below the threshold of the given similarity indicates injection of messages at the related time-window. Subsequently, the accuracy of the method in detecting injection of messages onto the CAN bus is computed. The best thresholds that we identified are $0.87$ for both the cosine similarity and Person correlation. Section~\ref{ssec:threasholdevaluate} discusses the results of the evaluation.

\begin{table}[tbp]
  \begin{center}
    \caption{Normality Test for metric variables (Shapiro-Wilk).}
    \label{tab:normality-test}
 \begin{tabular}{p{1.0in}p{0.9in}p{1.0in} }
 \hline \rowcolor{Gray} 
Data	&Cosine Similarity	&Pearson Similarity		\\\hline 
Normal 	&$1.17^{-5}$	&$1.12^{-5}$	 \\\hline
RPM	&$7.48^{-4}$	&$1.79^{-3}$		 \\\hline
Speed	&$9.995^{-12}$	&$9.899^{-12}$	 \\\hline

\hline
 
\end{tabular}
 \end{center}
\end{table}

\begin{figure*}[btp!]
    \centering
    \begin{subfigure}[t]{0.4\textwidth}
        \centering
        \includegraphics[width=1.0\linewidth]{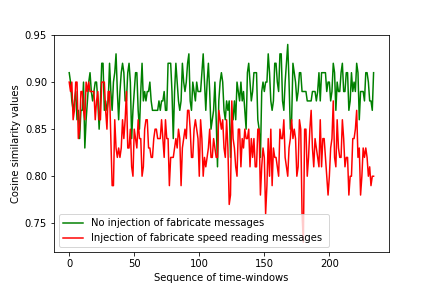}

 %       \label{speed }
    \end{subfigure}%
    ~
      \begin{subfigure}[t]{0.4\textwidth}
        \centering
        \includegraphics[width=1.0\linewidth]{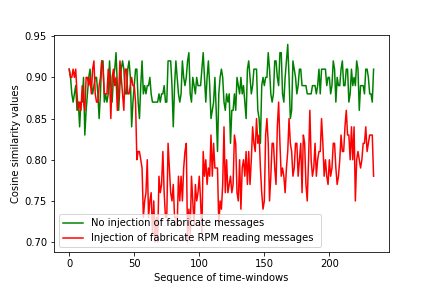}

  %      \label{rpm}
    \end{subfigure}%   
    \caption{Impact of the injection of speed reading messages (left) and RPM reading messages (right) on the of cosine similarity values of the MSGs of successive  time-windows; the cosine-similarity is higher without the injection of speed and RPM reading messages.}
    \label{Fig:CosinSimtimeseries}
\end{figure*}

\begin{figure*}[btp]
    \centering
    \begin{subfigure}[t]{0.4\textwidth}
        \centering
        \includegraphics[width=1.0\linewidth]{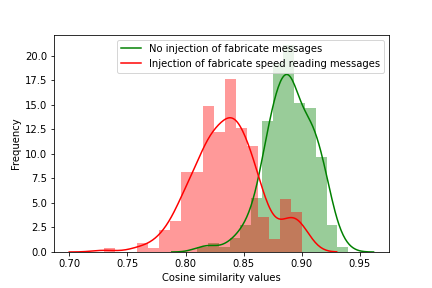}
        \label{CosinSimtimeseriesspeed}
    \end{subfigure}%\
    ~
     \begin{subfigure}[t]{0.4\textwidth}
        \centering
        \includegraphics[width=1.0\linewidth]{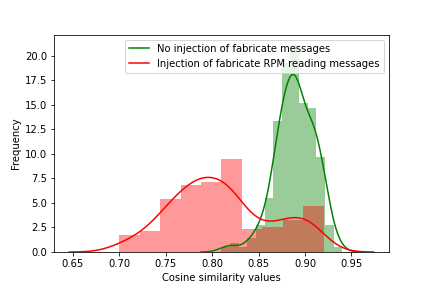}
        \label{CosinSimtimeseriesrpm}
    \end{subfigure}%
     \caption{Impact of the injection of speed reading messages (left) and RPM reading messages (right) on the distribution of the frequencies of the cosine similarity values of MSGs of successive time-windows.}
    \label{Fig:CosinSimHistogram}
\end{figure*}

\subsection{Evaluation method for the change point detection technique}

The \ac{CPD} method estimates the point of change of a population from a sample data. The method uses Markov-Chain Monte Carlo (MCMC) to sample the data and Bayesian inference to detect the point at which the mean of the population changes. The parameters of the model are the mean and the standard deviation of the population and the distributions of the data before the change point and after the change point. We choose the Normal distribution for both distributions. We tested the normality of cosine similarity values and Pearson correlation coefficients computed from the three datasets using the Shapiro-Wilk~\cite{ShWi1965}. The p-values of the t-tests are provided by Table~\ref{tab:normality-test}. The values indicate that all the datasets follow the Normal distribution. We used PYMC~\cite{PHF2010, HoGe2014} to evaluate the capability of the change point detection method to identify injection of CAN messages.

Using \ac{CPD} method to identify injection of messages from the similarity data, we compared the identified point of changes to the actual point of changes (i.e., the time we started injecting messages) for each of the datasets. In addition, we compare the \emph{strength of change} for each of the datasets, which is the proportion of the difference between the mean before the change point and after the change point to the average of the two means. We use threshold 1\% to interpret this strength of the change; a strength of change above 1\% implies there is a change and a strength of change below 1\% implies there is no change. Section~\ref{ssec:changepointevaluate} discusses the results of the evaluation.

\subsection{Evaluation method for the \ac{RNN}-\ac{LSTM}}

\begin{table}[tbp]
  \begin{center}
    \caption{RNN-LSTM model parameters.}
    \label{tab:LSTMconfiguration}
 \begin{tabular}{p{2.3 in}p{.8in}} \hline
 \rowcolor{Gray}\hline
 Parameters & Value \\
 \hline  
Input layer & 42 units \\
 \hline
 Second hidden layers & 12 units \\
 \hline
 Dense/Output layer & 1 units\\
 \hline
 Dropout rate & 20\% \\
 \hline
 Adam optimizer learning rate & 0.01 \\
 \hline
 Batch Size & 128 \\
 \hline 
 Number of epochs & 128\\
 \hline
 Ratio of the training dataset & 2/3\\
 \hline\hline
 
  \end{tabular}
 \end{center}
\end{table}

Initially, we developed prediction models using the \acf{RNN}-\acf{LSTM} method from the three raw-datasets described in Table~\ref{tab:listdatasets}. The models had low performance. To have datasets that have a balanced number of records wrt. injection/no injection of CAN messages, we constructed two datasets out the three row datasets as follows:
\begin{itemize}
    \item We appended the injection of speed reading messages onto the \ac{CAN} bus dataset to the no injection of fabricated messages dataset to form the \emph{constructed speed readings injection dataset}.
  \item We appended the injection of RPM reading messages onto the \ac{CAN} bus dataset to the no injection of fabricated messages dataset to form the \emph{constructed RPM readings injection dataset}.
\end{itemize}

We used \ac{LSTM}-\ac{RNN}, as implemented in package Keras~\cite{chollet2015keras}, to predict injection of messages in the CAN bus. Table~\ref{tab:LSTMconfiguration} lists the parameters that were used in the study. The sequences of \ac{MSG} similarities are fed onto the first/input layer of the LSTM model. The output of the input layer is passed on to the second layer. The output of the second hidden layer is then passed on to the output/dense layer, which maps the output values to binary values, representing the states injection of CAN messages and no injection of CAN messages. The experiment is performed for both dataset and the accuracy of the method is computed and reported in Section~\ref{ssec:LSTM-NNevaluate}.

\section{Results of the evaluation of the detection on messages injection using similarity of messages-sequences graphs}\label{sec:results}

This section describes the capability of \ac{MSG} similarity metric in conjunction with threshold, \ac{RNN}-\ac{LSTM}, and \ac{CPD} to detect injection on CAN messages. 

\subsection{Evaluation of the threshold technique}\label{ssec:threasholdevaluate}

This subsection describes the results of using cosine similarity and Pearson correlation of \acp{MSG} and thresholds to detect injection of CAN messages.

\begin{figure*}[tb]
    \centering
    \begin{subfigure}[b]{0.4\textwidth}
        \centering
        \includegraphics[width=1.0\linewidth]{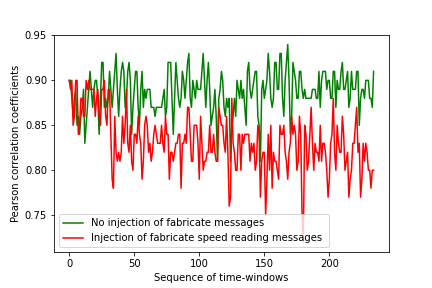}
    \end{subfigure}%
    ~
         \begin{subfigure}[b]{0.4\textwidth}
        \centering
        \includegraphics[width=1.0\linewidth]{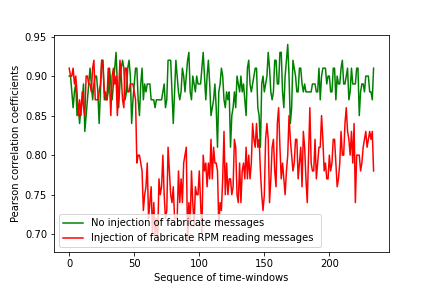}
    \end{subfigure}%

    \caption{Impact of the injection of speed reading messages (left) and RPM reading messages (right) on Pearson correlation coefficients of MSGs of successive  time-windows; the Pearson correlation coefficients are higher without the injection of speed and RPM reading messages.}
    \label{Fig:Pearsontimeseries}
\end{figure*}
\begin{figure*}[tb]
    \centering
    
    \begin{subfigure}[b]{0.4\textwidth}
        \centering
        \includegraphics[width=1.0\linewidth]{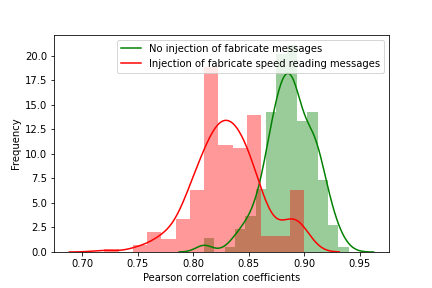}
    \end{subfigure}%
    ~
    \begin{subfigure}[b]{0.4\textwidth}
        \centering
        \includegraphics[width=1.0\linewidth]{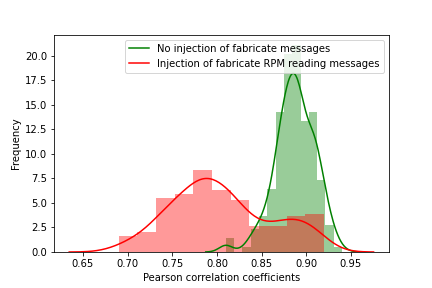}
    \end{subfigure}%
    \caption{Impact of injection of speed reading messages (left) and RPM reading messages (right) on the distribution of the frequencies of the Pearson correlation coefficients of MSGs of successive time-windows.}
    \label{Fig:PearsonHistogram}
\end{figure*}

\noindent{\bf Cosine Similarity Method.} Figure~\ref{Fig:CosinSimtimeseries} and Figure~\ref{Fig:CosinSimHistogram} show respectively the sequence of cosine similarity values and the distribution of the cosine similarity values computed from the three datasets. We observe from Figure~\ref{Fig:CosinSimtimeseries} that the plot of the sequence cosine similarity values extracted from the no injection of fabricated messages dataset is higher than the sequence of the cosine similarity values extracted from the injection of fabricated reading messages dataset. 

We observe from Figure~\ref{Fig:CosinSimHistogram} that the distribution of the cosine similarity values extracted from the no injection of fabricated messages dataset is different from the distribution of the cosine similarity values extracted from the injection of fabricated reading messages dataset in red color. The t-test confirms that injection of fabricated messages onto the CAN bus impacts the sequence of messages exchanged in the CAN bus. The t-test's p-value for the difference of the mean of the cosine similarity values extracted from the injection of speed reading messages dataset and the mean of the cosine similarity values extracted from the no injection of fabricated messages dataset is $1.12 e-74$ and the t-test's p-value for the difference of the mean of the cosine similarity values extracted from the injection of RPM reading messages dataset and the mean of the cosine similarity values extracted from the no injection of fabricated messages dataset is $2.61 e-69$. 

\noindent{\bf Pearson Correlation Method.} Figure~\ref{Fig:Pearsontimeseries} shows the sequence of correlation coefficients extracted from the three datasets over time, and Figure~\ref{Fig:PearsonHistogram} shows the distribution of correlation coefficients extracted from the three datasets. 

We observe from Figure~\ref{Fig:Pearsontimeseries} that the plot of the sequence Pearson  correlation coefficients extracted from the no injection of fabricated messages dataset is higher than the sequence of the Pearson  correlation coefficients extracted from the injection of fabricated reading messages dataset. We observe from Figure~\ref{Fig:PearsonHistogram} that the distribution of the Pearson  correlation coefficients extracted from the no injection of fabricated messages dataset is different from the distribution of the Pearson  correlation coefficients extracted from the injection of fabricated reading messages dataset. The t-test confirms that injection of fabricated messages onto the CAN bus impacts the sequence of messages exchanged in the CAN bus. 

 \begin{table}[tbp]
\caption{Performance of the three similarity metrics when using the injection of RPM readings dataset}
\label{tab:PerformanceRPMInjection}
\centering
\begin{tabular}{p{1.1in} p{.5in}p{.7in}p{.5in}}
\hline
\rowcolor{Gray}\hline
  Metrics & Accuracy (\%) & False Positive rate (\%) & Threshold\\\hline
\hline
Cosine similarity & 96.65 & 3.34 & 0.87 \\\hline
Pearson correlation & 97.32 & 2.67 & 0.87 \\ \hline
\hline
\end{tabular}
\end{table}

 \begin{table}[tbp]
\caption{Performance of the similarity metrics when using the injection of speed reading messages dataset.}
\label{tab:PerformanceSpeedinjection}
\centering
\begin{tabular}{p{1.1in} p{.5in}p{.7in}p{.5in}}
\rowcolor{Gray}\hline
  Metrics & Accuracy (\%) & False Positive rate (\%) & Threshold\\
\hline
Cosine similarity &89.20 &10.80 &0.87 \\\hline
Pearson correlation & 90.57 & 9.43 & 0.87 \\\hline
\hline
\end{tabular}
\end{table}

 \begin{table}[tbp]
\caption{Performance of the similarity metrics when using the no injection of fabricated RPM reading messages dataset.}
\label{tab:PerformanceNormal}
\centering
\begin{tabular}{p{1.1in} p{.5in}p{.7in}p{.5in}}
\hline
\rowcolor{Gray}
  Metric &Accuracy (\%) & False Positive rate (\%) & Threshold\\
\hline
Cosine similarity &89.4 &10.6 &0.87 \\\hline
Pearson correlation & 86.8 & 13.20 & 0.87 \\
\hline
\end{tabular}
\end{table}
The t-test's p-value for the difference of the mean of the Pearson  correlation coefficients extracted from the injection of speed reading messages dataset and the mean of the cosine similarity values extracted from the no injection of fabricated messages dataset is $9.466e-75$ and the t-test's p-value for the difference of the mean of the cosine similarity values extracted from the the injection of RPM reading messages dataset and the mean of the cosine similarity values extracted from the no injection of fabricated messages dataset is $1.6e-70$. 

\begin{figure*}[tbp]
    \centering
   \begin{subfigure}[t]{0.3\textwidth}
    \includegraphics[width=1.0\linewidth]{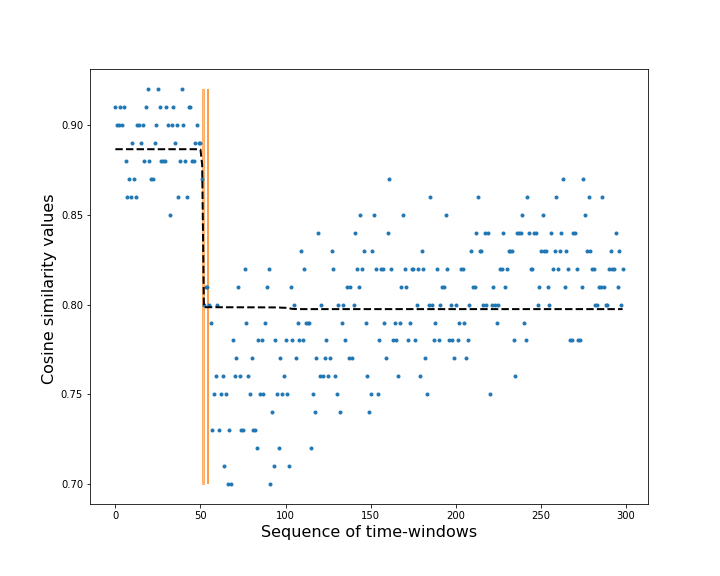}
    \centering
    %\caption{Change point detection for RPM readings dataset}
    \label{fig:cosinerpmchange}
    \end{subfigure}%\
    ~
   \begin{subfigure}[t]{0.3\textwidth}
    \includegraphics[width=1.0\linewidth]{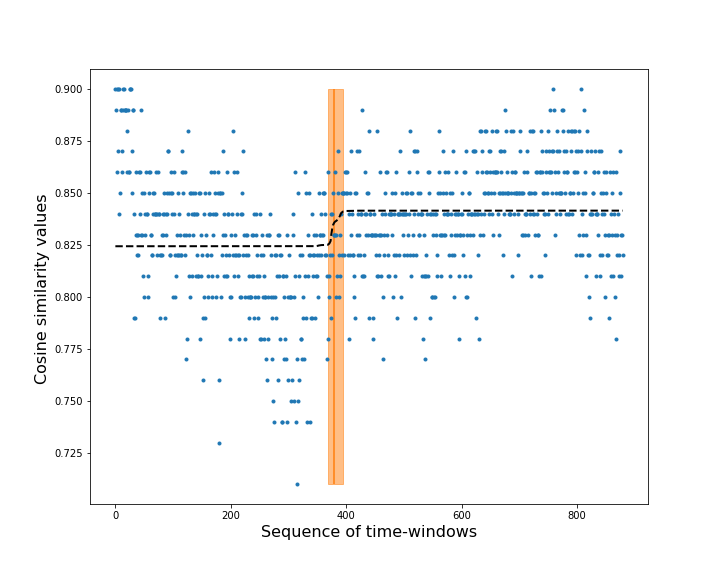}
    \centering
    %\caption{Change point detection for speed readings dataset}
    \label{fig:cosinespeedchange}
    \end{subfigure}%
    ~
    \begin{subfigure}[t]{0.3\textwidth}
    \includegraphics[width=1.0\linewidth]{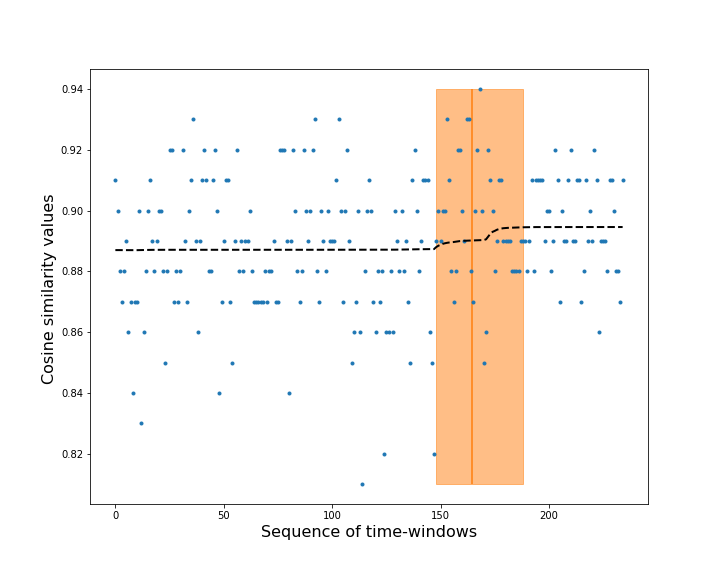}
    \centering
    %\caption{Change point detection for the no injection of fabricated messages dataset}
    \label{fig:cosinenormalchange}
    \end{subfigure}%\
    \caption{Detection of point of change of successive \acp{MSG} from the Cosine Similarity when RPM reading messages injected (left), speed reading messages injected (center), and normal operation (right).}
    \label{fig:cosinesimilaritytimeseries}
\end{figure*}

\begin{figure*}[tbp]
    \centering
     \begin{subfigure}[t]{0.3\textwidth}
        \centering
     \includegraphics[width=1.0\linewidth]{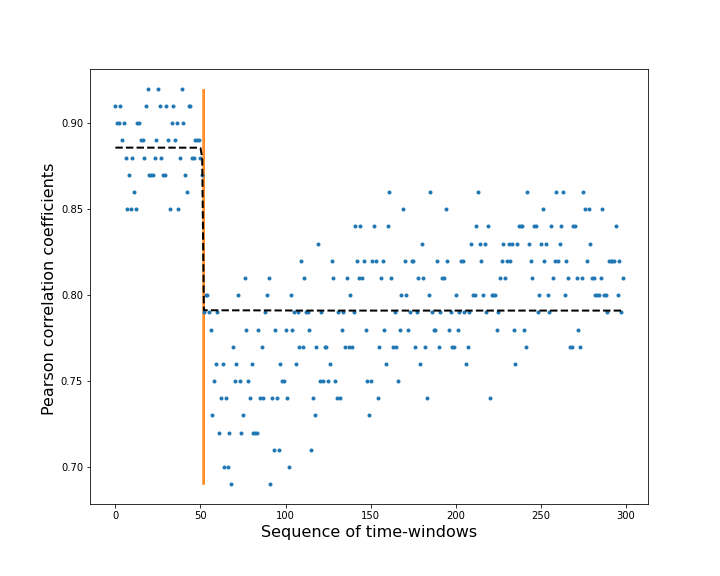}
    \centering
    %\caption{Change point detection from injection of RPM reading messages dataset}
    \label{fig:pearsonRPMChange}
    \end{subfigure}%
    ~
     \begin{subfigure}[t]{0.3\textwidth}
        \centering
     \includegraphics[width=1.0\linewidth]{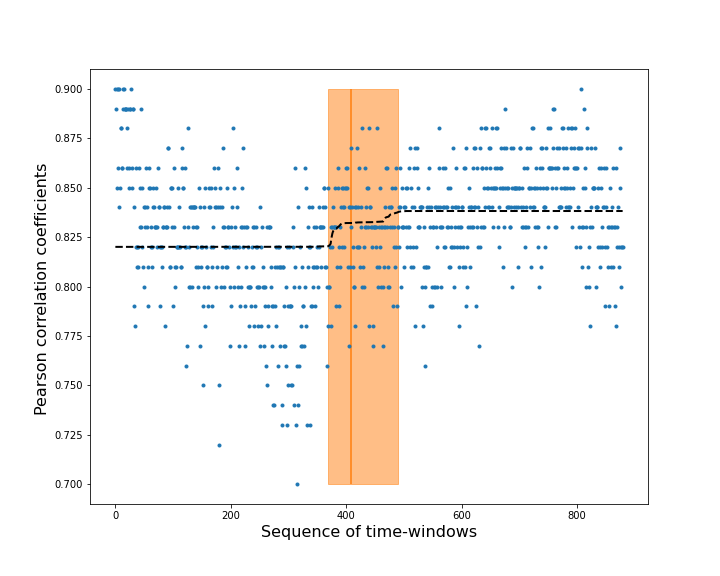}
    \centering
    %\caption{Change point detection for injection of speed reading messages dataset}
    \label{fig:pearsonspeedchange}
    \end{subfigure}%
      ~
     \begin{subfigure}[t]{0.3\textwidth}
        \centering
     \includegraphics[width=1.0\linewidth]{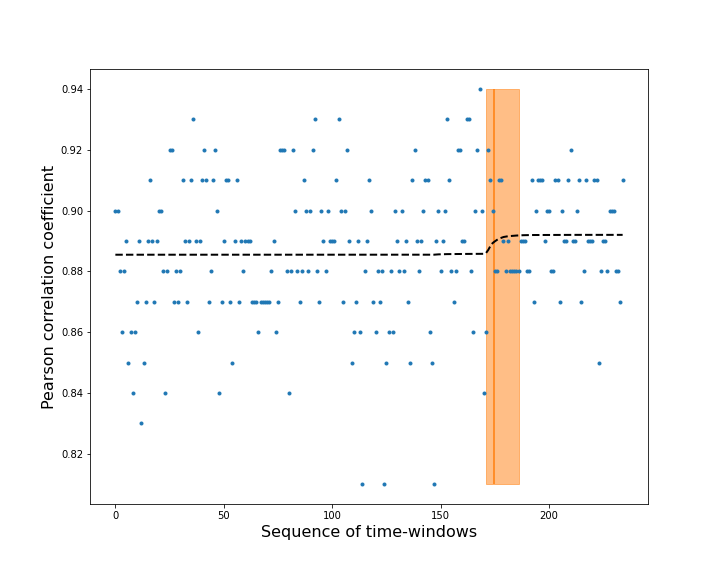}
    \centering
   % \caption{Change point detection from normal dataset}
    \label{fig:Pearsonnormalchange}
    \end{subfigure}%  
    \caption{Detection of point of change of successive \acp{MSG} from the Pearson correlation coefficients when RPM injected (left), speed message injection (center), and normal operation (right).}
    \label{fig:pearsoncorrelationtimeseries}
\end{figure*}

\noindent{\bf Accuracy of the Threshold Method.} Table~\ref{tab:PerformanceRPMInjection} and Table~\ref{tab:PerformanceSpeedinjection} summarize the accuracy of the threshold-based method in detecting injection of messages using the two similarity metrics applied to successive \acp{MSG} when using the three datasets. Table~\ref{tab:PerformanceNormal} shows also that the cosine similarity and Pearson correlation have low false-positive rates for the case of no injection of fabricated messages dataset. We conclude that cosine similarity and Pearson correlation exhibit excellent accuracy in detecting injection of speed and RPM reading messages.

\subsection{Change point detection method}\label{ssec:changepointevaluate}

This subsection describes the capability of the \ac{CPD} method in conjunction with the cosine similarity and Pearson correlation of \acp{MSG} to detect injection of messages onto the CAN bus.

 \begin{table}[tbp]
  \begin{center}
    \caption{Comparing the identified point of changes to the approximate actual point of change. (We set the point of change as the mean of the estimated point of change ranges with 94\%.) }
    \label{tab:change_point_pred}
 \begin{tabular}{p{1.2 in}p{.6in}p{.6in}p{.4in} }
 \hline \rowcolor{Gray} 
 Data&	Cosine similarity	&Pearson similarity	& Actual POC\\ \hline
Injection of RPM reading messages	&66	&54	& 34\\\hline
Injection of speed reading messages	&391	&468		&53	\\\hline
 No injection of fabricated messages  	&169	&174	&0	\\\hline

\end{tabular}
 \end{center}
\end{table}

\begin{table}[tbp]
  \begin{center}
    \caption{Ratio of the difference between the mean before the change point and after the change to the average of the two means (in percentage).}
    \label{tab:differnce_in_mean}

 \begin{tabular}{p{1.4 in}p{.8in}p{.8in}} 
 \hline \rowcolor{Gray} 
 Data&	Cosine Similarity	&Pearson Similarity		\\ \hline
Injection of RPM reading messages	&8.65	&10.96	\\\hline
Injection of speed reading messages	&2.07	&2.31	\\\hline
  No injection of fabricated messages 	&0.81	&18.12\\\hline

\end{tabular}

 \end{center}
\end{table}
 
Table~\ref{tab:change_point_pred} and shows the identified point of changes and the actual point of changes and Table~\ref{tab:differnce_in_mean} provides the strength of changes computed from the three datasets for both similarity techniques. Discussions on the capability of the two similarity metrics follow. 

\vspace{.2in}
\noindent{\bf Cosine similarity.} Figure~\ref{fig:cosinesimilaritytimeseries} shows the identified change points from the cosine similarity sequences computed from the injection of RPM reading messages, injection of speed reading messages and no injection of fabricated messages datasets. The change points are at windows 66, 391, and 169 for resp. The injection of RPM readings, injection of speed readings and no injection of fabricated messages datasets. We observe that the method detects quickly the change for the case of injection of RPM reading messages dataset (after 32 time-windows) but with significant delay for the case of injection of speed reading dataset dataset. 

We also observe that the method wrongly detects a change for the case of no injection of fabricated messages dataset. We observe, however, that the strength of change is 8.65\% for the injection of RPM readings, 2.07\% for the case of injection of speed readings dataset, and 0.81\% for the case of no injection of fabricated messages dataset. Therefore, the strength of the change metric indicates that the method detects change for the case of injection of RPM readings and injection of speed readings and does not detect change for the case of no injection of fabricated messages dataset. 

\begin{table*}[tbp]
\caption{State-of-the-art \ac{IDS} vs. our IDS solution. }
\label{tab:compareperformance}
\centering
\begin{tabular}{p{1.0in} p{3.5in}p{.6in}p{.6in}p{.5in}}
\rowcolor{Gray}
\hline
 IDS solution & Description & ML type & Knowledge of CAN ID& Accuracy\\\hline

IDS Supplier 1~\cite{SGL2019} &
 Detection of cyclic messages injection provided by IDS  Supplier 1 &Sup. &yes&1.0\\\hline
IDS Supplier 2~\cite{SGL2019} 
 &Detection of cyclic messages injection provided by IDS Supplier 2 &Sup. &yes &0.83\\\hline
Previous work~\cite{9076852}&Detection of injection of fabricated messages using HMM &Unsup.&yes&0.80\\\hline
\hline
Threshold-MSGs &Detection of injection of fabricated messages using \ac{MSG}, Pearson correlation similarity with threshold &Unsup.&no& $0.97$ \\ \hline
\ac{LSTM}-MSGs &Detection of injection of fabricated messages using \ac{MSG}, Pearson correlation similarity with \ac{LSTM}-\ac{RNN} &Sup.&no& 0.98\\ \hline
 \multicolumn{4}{l}{Note:}\\
  \multicolumn{5}{l}{(*) LSTM stands for \acl{LSTM}, \acf{RNN}, MSG stands for \acl{MSG}, Sup. is for}\\
   \multicolumn{4}{l}{\hspace{.2in} supervised ML technique and Unsup. is for unsupervised ML.}\\

\end{tabular}
\end{table*}

\begin{table}[tbp]
  \begin{center}
    \caption{Accuracy of LSTM-RNN in predicting injection of CAN messages.}
    \label{tab:PerformanceLSTM}
 \begin{tabular}{p{.35in}p{.6in}p{.53in}p{.6in}p{.53in} }
 \hline \rowcolor{Gray} \hline
 &\multicolumn{2}{p{1.3in}}{Constructed injection of speed reading messages dataset}&\multicolumn{2}{p{1.3in}}{Constructed injection of RPM reading messages dataset}\\\hline \rowcolor{lightgray} 
 Window	size &Cosine similarity &Pearson correlation  &Cosine similarity &Pearson correlation \\ \hline
100	& 96.93	&96.93& 97.32	&96.8		\\\hline
200&98.45	&98.45 &79.35	&55		\\ \hline
300	&73.43	&73.43 & 59.01&	88.52 		\\\hline
400	&93.75	&96.9&88.89&	88.89	\\\hline
500	&80.26	&100	&92.23&	97.1		\\\hline
600	&96.82	&96.8&37.93&	37.9		\\ \hline
700	&98.145	&98.145	&92         &92	\\\hline
800	&93.5	&93.5&90.48&	85.7		\\\hline
900	&97.5	&97.56&	83.33&	94.44		\\\hline
1000	&97.73	&97.3	&100&100	\\ \hline
\hline
  \end{tabular}
 \end{center}
\end{table}

\noindent{\bf Pearson correlation.} Figure~\ref{fig:pearsoncorrelationtimeseries} shows the identified change points from the Pearson correlation coefficients sequences computed from the injection of RPM reading messages, injection of speed reading messages, and no injection of fabricated messages datasets. The change points are at windows 54, 468, and 174 for resp. The injection of RPM reading messages, injection of speed reading messages and no injection of fabricated messages datasets. We observe that the change was quickly detected for the case of injection of RPM readings dataset (after 20 time-windows) but with a significant delay for the case of injection of speed readings dataset.

We also observe that the method wrongly detects a change for the no injection of fabricated messages dataset. We observe, however, that the strength of change is 10.96\% for the injection of RPM reading messages dataset, 2.31\% for the case of injection of speed reading messages dataset, and 0.75\% for the case of the no injection of fabricated message dataset. Therefore, the strength of the change metric indicates that the method detects change for the case of injection of RPM reading messages and injection of speed reading messages and does not detect change for the case of no injection of fabricated message dataset. 

\noindent{\bf Summary.} We observe that cosine similarity and Pearson correlation detect injection of RPM messages quickly but slow to detect injection of speed messages. In addition, we observe that change strength allows detecting message injection.

\subsection {LSTM-RNN prediction method}\label{ssec:LSTM-NNevaluate}

We discuss in the following the performance of \ac{LSTM}-\ac{RNN} in conjunction with the similarities of \acp{MSG} in predicting injection of messages onto the CAN bus.

Table~\ref{tab:PerformanceLSTM} provides the accuracy of the \ac{LSTM}-\ac{RNN} method in predicting injection of \ac{CAN} messages from a constructed injection of speed and RPM readings datasets considering different window sizes. We observe that the window size impacts the performance of LSTM-RNN in detecting injection of \ac{CAN} messages. The results show that 
\begin{enumerate}
    \item the accuracy for the cosine similarity varies between 73.43\% and 98.45\% for the case of constructed injection of speed reading messages dataset and  between 37.93\% and 100\% for the case of constructed injection of RPM reading messages dataset;
    \item the accuracy for the Pearson correlation varies between 73.43\% and 100\% for the case of constructed injection of speed reading messages dataset and between 37.9\% and 100\% for the case of constructed injection of speed reading messages dataset.
\end{enumerate} 

We conclude that \ac{LSTM}-\ac{RNN} exhibits excellent accuracy in predicting injection of \ac{CAN} messages onto the CAN bus from the constructed injection of speed and RPM  readings messages dataset when using cosine similarity and Pearson correlation to compute the similarities of the consecutive \acp{MSG}. 

\subsection{Impact of the results}\label{ssec:discussion}

%They showed that they achieve 100\% accuracy for a given manufacturer, referred to as Supplier 1, and 83\% accuracy from Supplier 2 using supervised (Neural Net) machine learning approach as illustrated in Table~\ref{tab:compareperformance}. The NN-based IDS of their study was evaluated on three vehicle makes (Supplier 1) with 100\%, however, the training dataset depends on the knowledge of the association of CAN IDs to the corresponding \acp{ECU} for each of the vehicle make, furthermore, the supervised ML technique employed requires extensive time to develop high-performance IDS models from labeled data. 

%In our previous study \cite{9076852}, we attempted unsupervised machine learning technique using Hidden Markov Model (HMM) and achieved 80\% accuracy. In this work, we employed a combination of Pearson similarity correlation and Threshold and improved to 97\% accuracy, and with LSTM-RNN techniques we reached 98\% accuracy. Although our findings didn't achieve higher accuracy, we showed a detection and prediction scheme without the knowledge of the propriety CAN bus ID information. Note that we used in this study three datasets related to one driver and one vehicle. We consider experimental evaluation for a set of vehicles and drivers with improved accuracy in our future work.

Although there are several machine learning-based message injection attack detection studies in the literature, we found, to the best of our knowledge, Stachowski et al.~\cite{SGL2019} is the closest study related to our work since they use datasets from in-motion vehicles. Table~\ref{tab:compareperformance} compares the IDS solutions that we developed with the two pre-commercials \acp{IDS} (Supplier 1 and Supplier 2) that were evaluated by Stachowski et al.~\cite{SGL2019} and the IDS that we proposed earlier~\cite{9076852}. One of the NN-based IDS that was evaluated by Stachowski et al. on three vehicles makes (each was tested for several hours) showed complete accuracy (1.0) but requires knowledge of the association of CAN IDs to the corresponding \acp{ECU} for each of the vehicle make and requires the use of a pre-trained prediction model--uses a supervised ML technique.

Our proposed IDS solution does not require the knowledge of the association of CAN IDs to the corresponding \acp{ECU} of the vehicle and shows good accuracy, up to 0.98, when a supervised ML (RNN-LSTM) is used, and with high accuracy when an unsupervised method (threshold technique) is used, up to 0.97. This shows that our scheme is a practical \emph{generic} IDS solution as it does not use proprietary information from the car manufacturers. Additionally, the \ac{CPD}-\acp{MSG} method showed a quick detection of injection of fabricated messages as discussed in the previous section. Note that we used in this study three datasets related to one driver and one vehicle. We consider experimental evaluation for a set of vehicles and drivers in our future work.

\section{Conclusions}\label{sec:conclusions}

The number of cyber-attacks on connected vehicles is increasing. The research community has proposed anomaly-based \acf{IDS} solutions to address this problem. The main advantage of this solutions is that it does not require modification to the CAN protocol. Existing \acp{IDS} either use the entropy of messages or require the knowledge of the IDs of the different sensors/actuators of the test vehicle. This paper investigates the use of \acf{MSG}, which models the sequences of CAN messages in a time window to detect message injection. The study found that the cosine similarity and Pearson correlation of the sequence of \acp{MSG} are effective in detecting injection of RPM and speed messages with 98.45\% accuracy. 

\section{Acknowledgement}

The authors thank Bhagath-Kuma Veerannagari for developing some of the functions used in the research. This research is partly funded by Iowa State University's Regents Innovation Fund (RIF).

%\afterpage{\clearpage}

\bibliographystyle{ieeetr}
\bibliography{AttackDetection}

\begin{thebibliography}{10}

\bibitem{bosh1991}
R.~{Bosch GmbH}, ``Can specification v2.0.''
  \url{http://esd.cs.ucr.edu/webres/can20.pdf}, 1991.

\bibitem{4562233}
D.~Nilsson, P.~Phung, and U.~Larson, ``Vehicle {ECU} classification based on
  safety-security characteristics,'' in {\em Proc. IET Road Transport
  Information and Control}, (Manchester, UK), pp.~1--7, 2008.

\bibitem{Othmane2015}
L.~{ben Othmane}, H.~Weffers, M.~M. Mohamad, and M.~Wolf, {\em Wireless Sensor
  and Mobile Ad-Hoc Networks: Vehicular and Space Applications}, ch.~A Survey
  of Security and Privacy in Connected Vehicles, pp.~217--247.
\newblock Springer, 2015.

\bibitem{HHZ2018}
V.~H. Le, J.~Hartog, and N.~Zannone, ``Security and privacy for innovative
  automotive applications: A survey,'' {\em Computer Communications}, vol.~132,
  pp.~17 -- 41, 2018.

\bibitem{OABF2018}
L.~ben Othmane, V.~Alvarez, K.~Berner, M.~Fuhrmann, W.~Fuhrmann, A.~Guss, and
  T.~Hartsock, ``Demo: A low-cost fleet monitoring system,'' in {\em The Fourth
  IEEE Annual International Smart Cities Conference}, (Kansas City, MO), Sep.
  2018.

\bibitem{UpstreamAuto2020}
{Upstream Auto}, ``Upstream security's global automotive cybersecurity
  report.''
  \url{https://www.upstream.auto/upstream-security-global-automotive-cybersecurity-report-2020/},
  2020.
\newblock accessed on Feb 2020.

\bibitem{ValMill}
C.~Miller and C.~Valasek, ``Adventures in automotive networks and control
  units.'' \url{http://illmatics.com/carhacking.pdf}, 2013.

\bibitem{Brandom}
R.~Brandom, ``The scariest thing about the chrysler hack is how hard it was to
  patch,'' Jul 2015.

\bibitem{Golson}
J.~Golson, ``Car hackers demonstrate wireless attack on tesla model s.''
  \url{https://www.theverge.com/2016/9/19/12985120/tesla-model-s-hack-vulnerability\-keen-labs},
  Sep 2016.

\bibitem{9076852}
L.~{Ben Othmane}, L.~{Dhulipala}, N.~{Multari}, and M.~{Govindarasu}, ``On the
  performance of detecting injection of fabricated messages into the can bus,''
  {\em IEEE Transactions on Dependable and Secure Computing}, pp.~1--1, 2020.

\bibitem{SGL2019}
S.~Stachowski, R.~Gaynier, and D.~J. LeBlanc, ``An assessment method for
  automotive intrusion detection system performance.''
  \url{https://rosap.ntl.bts.gov/view/dot/41006}, April 2019.

\bibitem{Embedded-system}
T.~G. Frank~Vahid, ``Embedded system design: A unified hardware/software
  approach.'' \url{http://dsp-book.narod.ru/ESDUA.pdf}, 1999.
\newblock Accessed on 05/02/2020.

\bibitem{OBDII}
B.~Electronics, ``Obd-ii - on-board diagnostic system.''
  \url{http://www.obdii.com/connector.html}.
\newblock accessed on Jan. 2019.

\bibitem{DBLP:journals/corr/abs-1802-01725}
O.~Avatefipour and H.~Malik, ``State-of-the-art survey on in-vehicle network
  communication (can-bus) security and vulnerabilities,'' {\em CoRR},
  vol.~abs/1802.01725, 2018.

\bibitem{article1}
S.~Hartzell, C.~Stubel, and T.~Bonaci, ``Security analysis of an automobile
  controller area network bus,'' {\em IEEE Potentials}, vol.~39, pp.~19--24, 05
  2020.

\bibitem{article}
M.~Wolf, A.~Weimerskirch, and T.~Wollingerand, ``State of the art: Embedding
  security in vehicles,'' {\em EURASIP Journal on Embedded Systems}, vol.~2007,
  pp.~1--16, 01 2007.

\bibitem{25}
C.~Valasek and C.~Miller, ``Adventures in automotive networks and control
  units.''
  \url{https://ioactive.com/pdfs/IOActive_Adventures_in_Automotive_Networks_and_Control_Units.pdf},
  2015.
\newblock Accessed on 01/02/2020.

\bibitem{golson2016}
G.~Golson, ``Car hackers demonstrate wireless attack on tesla model s.'' \url
  {https://www.theverge.com/2016/9/19/12985120/tesla-model-s-hack-vulnerability-keen-labs},
  2016.
\newblock Accessed on 03/02/2020.

\bibitem{green2020}
A.~Greenburg, ``This bluetooth attack can steal a tesla model x in minutes.''
  \url {https://www.wired.com/story/tesla-model-x-hack-bluetooth/}.
\newblock Accessed on 03/02/2020.

\bibitem{5504804}
K.~{Koscher}, A.~{Czeskis}, F.~{Roesner}, S.~{Patel}, T.~{Kohno}and
  S.~{Checkoway}, D.~{McCoy}, B.~{Kantor}, D.~{Anderson}, H.~{Shacham}, and
  S.~{Savage}, ``Experimental security analysis of a modern automobile,'' in
  {\em Proc. 2010 IEEE Symposium on Security and Privacy (SP)}, (Los Alamitos,
  CA, USA), pp.~447--462, may 2010.

\bibitem{OFRB2014}
L.~{ben Othmane}, R.~Fernando, R.~Ranchal, B.~Bhargava, and E.~Bodden,
  ``Likelihood of threats to connected vehicles,'' {\em International Journal
  of Next-generation Computing (IJNGC)}, vol.~5, pp.~290--303, Nov. 2014.

\bibitem{7934878}
W.~A. {Farag}, ``Cantrack: Enhancing automotive can bus security using
  intuitive encryption algorithms,'' in {\em 2017 7th International Conference
  on Modeling, Simulation, and Applied Optimization (ICMSAO)}, pp.~1--5, 2017.

\bibitem{28}
H.~Ueda, T.~M. Ryo Kurachiand Hiroaki~Takada, M.~Inoue, , and S.~Horihata,
  ``Security authentication system for in-vehicle network,'' {\em SEI technical
  review}, vol.~81, pp.~5--9, 2015.

\bibitem{inproceedings}
A.~Herrewege, D.~Singelée, and I.~Verbauwhede, ``Canauth - a simple, backward
  compatible broadcast authentication protocol for can bus,'' in {\em Proc.
  ECRYPT Workshop on Lightweight Cryptography}, (Dresden, Germany), p.~7,
  Proc.In Embedded Security in Cars 9th, 01 2011.

\bibitem{10.1007/978-3-642-35404-5_15}
B.~Groza, S.~Murvay, A.~van Herrewege, and I.~Verbauwhede, ``Libra-can: A
  lightweight broadcast authentication protocol for controller area networks,''
  in {\em Proc. Cryptology and Network Security}, (Berlin, Heidelberg"),
  pp.~185--200, Springer Berlin Heidelberg, 2012.

\bibitem{inproceedings2}
R.~Kurachi, Y.~Matsubara, H.~Takada, N.~Adachi, Y.~Miyashita, and S.~Horihata,
  ``Cacan - centralized authentication system in can,'' in {\em Proc. Embedded
  Security in Cars(escar) Europe}, 11 2014.

\bibitem{inproceedings3}
R.~Kurachi, T.~Pyun, S.~Honda, H.~Takada, H.~Ueda, and S.~Horihata, ``Can
  disabler: Hardware-based prevention method of unauthorized transmission in
  can and can-fd networks,'' in {\em Proc. Embedded Security in Cars(escar)},
  06 2016.

\bibitem{article3}
M.~Bozdal, M.~Samie, S.~Aslam, and I.~Jennions, ``Evaluation of can bus
  security challenges,'' {\em Sensors}, vol.~20, pp.~16--17, 04 2020.

\bibitem{7420322}
A.~{Taylor}, N.~{Japkowicz}, and S.~{Leblanc}, ``Frequency-based anomaly
  detection for the automotive can bus,'' in {\em Proc. 2015 World Congress on
  Industrial Control Systems Security (WCICSS)}, pp.~45--49, 2015.

\bibitem{10.1371/journal.pone.0155781}
M.-J. Kang and J.-W. Kang, ``Intrusion detection system using deep neural
  network for in-vehicle network security,'' {\em PLOS ONE}, vol.~11,
  pp.~1--17, 06 2016.

\bibitem{Lokman2019}
S.-F. Lokman, A.~T. Othman, and M.-H. Abu-Bakar, ``Intrusion detection system
  for automotive controller area network (can) bus system: a review,'' {\em
  EURASIP Journal on Wireless Communications and Networking}, vol.~184, no.~1,
  2019.

\bibitem{8687274}
S.~{Lokman}, A.~T. {Bin Othman}, and M.~{Abu-Bakar}, ``Optimised structure of
  convolutional neural networks for controller area network classification,''
  in {\em 2018 14th International Conference on Natural Computation, Fuzzy
  Systems and Knowledge Discovery (ICNC-FSKD)}, pp.~475--481, July 2018.

\bibitem{9262960}
S.~{Longari}, D.~H.~N. {Valcarcel}, M.~{Zago}, M.~{Carminati}, and S.~{Zanero},
  ``Cannolo: An anomaly detection system based on lstm autoencoders for
  controller area network,'' {\em IEEE Transactions on Network and Service
  Management}, pp.~1--1, 2020.

\bibitem{SHH2018}
E.~{Seo}, H.~M. {Song}, and H.~K. {Kim}, ``Gids: Gan based intrusion detection
  system for in-vehicle network,'' in {\em Proc. 2018 16th Annual Conference on
  Privacy, Security and Trust (PST)}, pp.~1--6, Aug 2018.

\bibitem{8688625}
W.~{Wu}, R.~{Li}, G.~{Xie}, J.~{An}, Y.~{Bai}, J.~{Zhou}, and K.~{Li}, ``A
  survey of intrusion detection for in-vehicle networks,'' {\em IEEE
  Transactions on Intelligent Transportation Systems}, vol.~21, no.~3,
  pp.~919--933, 2020.

\bibitem{IRYM2020}
R.~{Islam}, R.~U.~D. {Refat}, S.~M. {Yerram}, and H.~{Malik}, ``Graph-based
  intrusion detection system for controller area networks,'' {\em IEEE
  Transactions on Intelligent Transportation Systems}, pp.~1--10, Oct, 2020.

\bibitem{ChSh2016}
K.-T. Cho and K.~G. Shin, ``Fingerprinting electronic control units for vehicle
  intrusion detection,'' in {\em Proc. of the 25th USENIX Conference on
  Security Symposium}, (Austin, TX, USA), pp.~911--927, 08 2016.

\bibitem{TLJ20168}
A.~Taylor, S.~Leblanc, and N.~Japkowicz, ``Anomaly detection in automobile
  control network data with long short-term memory networks,'' in {\em Proc.
  IEEE International Conference on Data Science and Advanced Analytics (DSAA)},
  (Montreal, Canada), pp.~130--139, Oct 2016.

\bibitem{LMKA2017}
M.~Levi, Y.~Allouche, and A.~Kontorovich, ``Advanced analytics for connected
  car cybersecurity,'' in {\em Proc. IEEE 87th Vehicular Technology Conference
  (VTC Spring)}, (Porto, Portugal), pp.~1--7, 06 2018.

\bibitem{9235336}
G.~{D’Angelo}, A.~{Castiglione}, and F.~{Palmieri}, ``A cluster-based
  multidimensional approach for detecting attacks on connected vehicles,'' {\em
  IEEE Internet of Things Journal}, pp.~1--1, 2020.

\bibitem{CROT2016}
D.~S. Cruzes and L.~ben Othmane, {\em Empirical Research for Software
  Security{:} Foundations and Experience}, ch.~Threats to Validity in Software
  Security Empirical Research, pp.~275--300.
\newblock Taylor \& Francis Group, LLC, 2017.

\bibitem{othmane2020b}
L.~{Ben othmane} and L.~Dhulipala, ``Injection of rpm and speed reading
  messages onto the can bus of a moving vehicle.''
  \url{https://dx.doi.org/10.21227/s1jy-h433}, 2020.
\newblock IEEE Dataport.

\bibitem{8640808}
C.~{Young}, J.~{Zambreno}, H.~{Olufowobi}, and G.~{Bloom}, ``Survey of
  automotive controller area network intrusion detection systems,'' {\em IEEE
  Design Test}, vol.~36, no.~6, pp.~48--55, 2019.

\bibitem{Neman2010}
M.~Newman, {\em Networks - An Introduction}.
\newblock Oxford University Press, 2010.

\bibitem{TJS2018}
T.~Andrew, J.~Bryans, and S.~Shaikh, ``Towards viable intrusion detection
  methods for the automotive controller area network,'' in {\em Proc. 2nd
  Computer Science in Cars Symposium - Future Challenges in Artificial
  Intelligence Security for Autonomous Vehicles (CSCS 2018)}, (Munich,
  Germany), Sep 2018.

\bibitem{279181}
Y.~{Bengio}, P.~{Simard}, and P.~{Frasconi}, ``Learning long-term dependencies
  with gradient descent is difficult,'' {\em IEEE Transactions on Neural
  Networks}, vol.~5, no.~2, pp.~157--166, 1994.

\bibitem{HoSc1997}
S.~Hochreiter and J.~Schmidhuber, ``Long short-term memory,'' {\em Neural
  Computation}, vol.~9, November 1997.

\bibitem{OEMG2019}
H.~Olufowobi, U.~Ezeobi, E.~Muhati, G.~Robinson, C.~Young, J.~Zambreno, and
  G.~Bloom, ``Anomaly detection approach using adaptive cumulative sum
  algorithm for controller area network,'' in {\em Proc. ACM Workshop on
  Automotive Cybersecurity}, AutoSec '19, (Richardson, USA), p.~25–30, 2019.

\bibitem{BaHa1993}
D.~Barry and J.~A. Hartigan, ``A bayesian analysis for change point problems,''
  {\em Journal of the American Statistical Association}, vol.~88, no.~421,
  pp.~309--319, 1993.

\bibitem{Gosset1908}
W.~S. Gosset, ``The probable error of a mean,'' {\em Biometrika}, vol.~6,
  no.~1, pp.~1--25, 1908.

\bibitem{ShWi1965}
S.~Shapiro and M.~Wilk, ``An analysis of variance test for normality (complete
  samples).,'' {\em Biometrika}, vol.~52, no.~3/4, p.~591–611, 1965.

\bibitem{PHF2010}
A.~Patil, D.~Huard, and C.~J. Fonnesbeck, ``Pymc: Bayesian stochastic modelling
  in python,'' {\em Journal of Statistical Software}, vol.~35, no.~4,
  pp.~1--81, 2010.

\bibitem{HoGe2014}
M.~D. Hoffman and A.~Gelman, ``The no-u-turn sampler: Adaptively setting path
  lengths in hamiltonian monte carlo,'' {\em Journal of Machine Learning
  Research}, vol.~15, no.~47, pp.~1593--1623, 2014.

\bibitem{chollet2015keras}
F.~Chollet {\em et~al.}, ``Keras.'' \url{https://github.com/fchollet/keras},
  2015.
\newblock Accessed on Feb. 2020.

\end{thebibliography}

\vspace{-0.6in}
\begin{IEEEbiography}[{\includegraphics[width=1in,height=1.25in,clip,keepaspectratio]{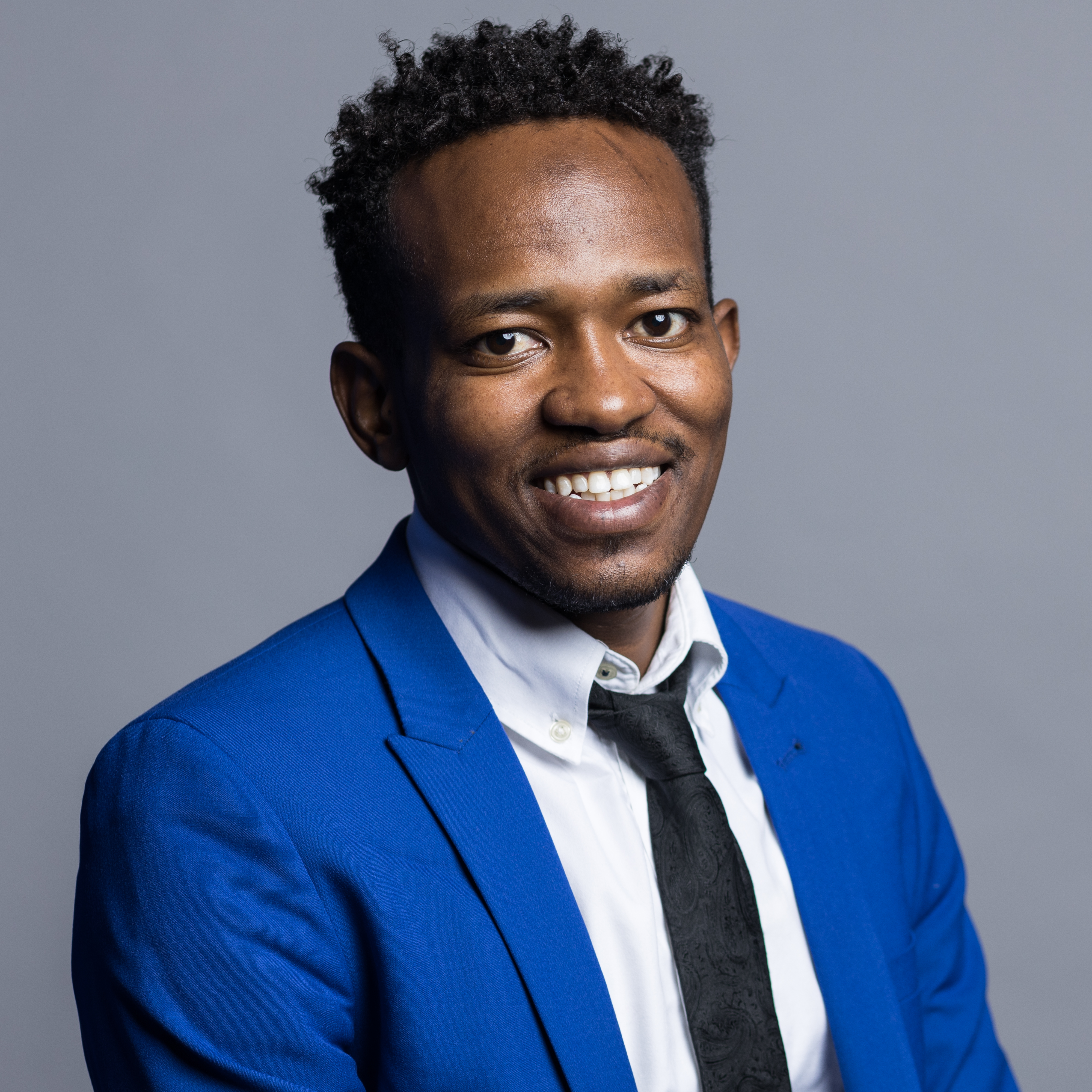}}]{Mubark Jedh}
   Mubark Jedh is currently a Ph.D student at Iowa State University. He received his B.S. from Northern Illinois University, USA in 2017, and M.S. from Iowa State University, USA in 2020.   
\end{IEEEbiography}

\vspace{-0.7in}
\begin{IEEEbiography}[{\includegraphics[width=1in,height=1.25in,clip,keepaspectratio]{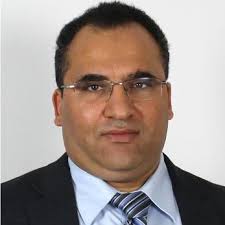}}]{Lotfi ben Othmane}
  Dr. Lotfi ben Othmane is Assistant Teaching Professor at the Department of Electrical and Computer Engineering and is leading the Engineering Secure Smart Cyber-Physical Systems Lab at Iowa State University, USA. Previously, he was a Research Scientist and then Head of the Secure Software Engineering department at Fraunhofer SIT, Germany. Lotfi received his Ph.D. from Western Michigan University (WMU), USA, in 2010; the M.S. in computer science from University of Sherbrooke, Canada, in 2000; and the B.S  in information systems from University of Economics and Management of Sfax, Tunisia, in 1995. He works currently on engineering secure cyber-physical systems. 
\end{IEEEbiography}

\vspace{-0.7in}
\begin{IEEEbiography}[{\includegraphics[width=1in,height=1.25in,clip,keepaspectratio]{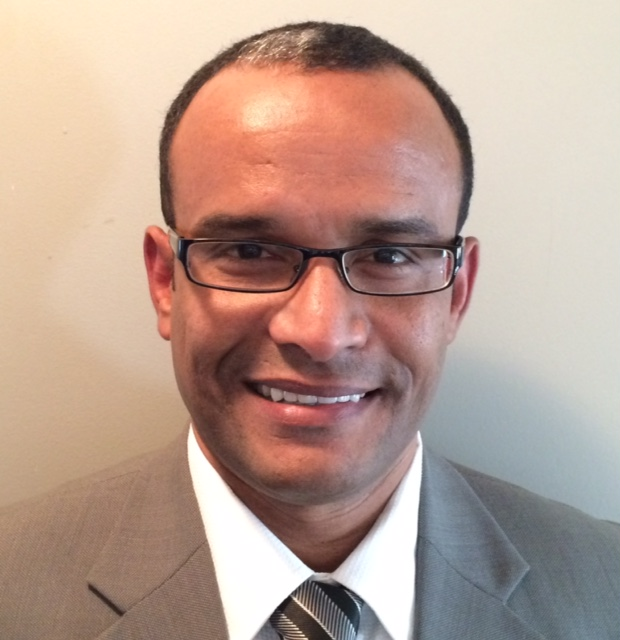}}]{Noor Ahmed, PhD.} is a Computer Scientist at the Air Force Research Laboratory in Rome, NY since 2003.  He holds a BSc (2002) from Utica College, MSc (2006) from Syracuse University, and PhD (2016) from Purdue University, all in Computer Science. His research focuses on; Security in Cloud Computing and SOA, Semantic Computing, Reliability and Resiliency in Distributed Systems with spacial emphasis on Moving Target Defense, Blockchain, and Cyber Physical Systems.
 
\end{IEEEbiography}
\vspace{-0.7in}

\begin{IEEEbiography}[{\includegraphics[width=1in, height=1.25in, clip,keepaspectratio]{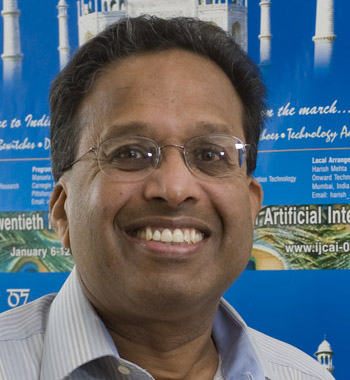}}]{Bharat Bhargava} is with the Department of Computer Science at Purdue University, West Lafayette, IN. He received his Ph.D. in Electrical Engineering from Purdue University. He is a fellow of the IEEE and IETE. (bbshail@purdue.edu).
\end{IEEEbiography}

\end{document}